\documentclass[12pt]{article}
\usepackage[breaklinks=true,linktocpage=true]{hyperref}
\usepackage{ascmac}
\usepackage{graphicx}
\usepackage{xcolor}
\usepackage{cite}

\usepackage{amsmath}
\usepackage{amssymb,amsthm}
\usepackage{amscd} 
\usepackage[all]{xy}
\usepackage{mathrsfs}
\numberwithin{equation}{section}
\usepackage{mathtools}

\usepackage{caption}
\usepackage[top=0.14\paperwidth,bottom=0.14\paperwidth,left=0.14\paperwidth,right=0.14\paperwidth]{geometry}

\newcommand{\link}{\,{\rm Link}\,}

\newcommand{\ig}{\includegraphics}

\newcommand{\vevs}[1]{\langle #1 \rangle}

\newcommand{\er}[1]{Eq.~\eqref{#1}}
\newcommand{\ers}[1]{Eqs.~\eqref{#1}}

\newcommand{\bb}{\mathbb}

\newcommand{\hph}{\hphantom}
\renewcommand{\b}{\bar}

\newcommand{\bs}{\boldsymbol}

\newcommand{\fr}{\frac}

\newcommand{\der}{\partial}

\renewcommand{\(}{\left(}
\renewcommand{\)}{\right)}

\newcommand{\wed}{\wedge}
\newcommand{\weds}{\wedge \cdots \wedge}
\newcommand{\bmx}{\left(\begin{matrix}}
\newcommand{\emx}{\end{matrix}\right)}

\allowdisplaybreaks[4]
\begin{document}
\renewcommand{\thefootnote}{\fnsymbol{footnote}}%
\begin{titlepage}
\hfill 
KEK-TH-2346, J-PARC-TH-0249
\vspace{-1em}
 \vspace{3em}
 \begin{center}%
  {\LARGE 
Global 4-group symmetry and 't Hooft anomalies
\\
in topological axion electrodynamics 
  \par
   }%
 \vspace{1.5em} 
 {\large
 Yoshimasa Hidaka,\footnote{hidaka@post.kek.jp}${}^{a,b,c}$
    Muneto Nitta,\footnote{nitta@phys-h.keio.ac.jp}${}^{d}$
and
    Ryo Yokokura\footnote{ryokokur@post.kek.jp}${}^{a,d}$
   \par
   }%
  \vspace{1em} 
${}^a${\small\it KEK Theory Center, Tsukuba 305-0801, Japan}
\par
${}^b$
{\small\it Graduate University for Advanced Studies (Sokendai), Tsukuba 305-0801, Japan}
\par
${}^c$
{\small\it RIKEN iTHEMS, RIKEN, Wako 351-0198, Japan}
\par
${}^d${\small\it Department of Physics \& Research and Education Center for Natural Sciences,
\par 
Keio University, Hiyoshi 4-1-1, Yokohama, Kanagawa 223-8521, Japan}
\par
  \vspace{1em} 
 \end{center}%
 \par
\vspace{1.5em}%
\begin{abstract}
We study higher-form global symmetries and a higher-group structure 
of a low-energy limit of $(3+1)$-dimensional 
axion electrodynamics in a gapped phase
described by a topological action.
We argue that the higher-form symmetries should have a 
semi-strict 4-group 
(3-crossed module) structure by consistency conditions of 
couplings of the topological action to background gauge fields 
for the higher-form symmetries.
We find possible 't Hooft anomalies for the 4-group global symmetry, 
and discuss physical consequences. 
\end{abstract}
\end{titlepage}
\setcounter{footnote}{0}%
\renewcommand{\thefootnote}{$*$\arabic{footnote}}%
\tableofcontents

\section{Introduction}

Axions are hypothetical pseudo-scalar bosons, and have been studied in
various contexts in modern physics, such as particle physics, cosmology,
string theory, hadron physics, and condensed matter physics.
In the context of particle physics, the axion, called the QCD axion, was
introduced as a candidate for the solution to the strong CP problem~\cite{Peccei:1977hh,Weinberg:1977ma,Wilczek:1977pj,Dine:1981rt,Zhitnitsky:1980tq,Kim:1979if,Shifman:1979if}.
Later the axion was considered to be a candidate for dark matter as well~\cite{Preskill:1982cy,Abbott:1982af,Dine:1982ah,Stecker:1982ws}.
The axion was also generalized to axion-like particles~\cite{Maiani:1986md,Raffelt:1987im}, 
which
are not necessarily to solve the strong CP problem, but they 
have
similar virtues to the QCD axion, as we will mention below.
Such axion-like particles can naturally arise as moduli fields in
4-dimensional effective theories of string theory~\cite{Witten:1985xb,Townsend:1993wy,Izquierdo:1993st,Harvey:2000yg,Svrcek:2006yi,Arvanitaki:2009fg}.
They have also been discussed as candidates for dark
matter~\cite{Masso:1995tw} 
or an inflaton that can cause inflation in the early universe~\cite{Freese:1990rb,Adams:1992bn}.
In the context of condensed matter physics, axions have been regarded as
quasi-particle excitations~\cite{Li:2009tca,Wang:2012bgb} 
or parameters that characterize topological
insulators~\cite{Wilczek:1987mv,Qi:2008ew,Essin:2008rq}
\footnote{See e.g., Refs.~\cite{Kim:1986ax,Dine:2000cj,Peccei:2006as,Kawasaki:2013ae,Marsh:2015xka}
and Refs.~\cite{Hasan:2010xy,Sekine:2020ixs} as reviews 
of axions in particle physics and condensed matter physics, respectively.}.

One of the virtues of the axions is a topological coupling to a photon.
Here, the ``topological coupling'' means that it does not depend on the
metric of the spacetime.
Such a coupling exists as a result of a chiral anomaly of Dirac fermions coupled to the axion and photon.
Thus, the coupling is stable against higher-order radiative corrections.
The axion-photon coupling plays important roles in the above
applications, such as a decay of the axions to photons,
magneto-electric responses, and so on~\cite{Fischler:1983sc,Sikivie:1984yz,Kaplan:1987kh,Manohar:1988gv,Wilczek:1987mv,Kogan:1992bq,Kogan:1993yw,Qi:2008ew,Essin:2008rq,Ferrer:2015iop,Yamamoto:2015maz,Ferrer:2016toh}.
The simplest model given by only an axion and a photon with a topological coupling is called the ``axion electrodynamics''~\cite{Wilczek:1987mv}.

There can be several phases of the axion electrodynamics 
according to the mass gaps of the axion and photon.
In particular, the phase in which both 
the axion and photon have mass gaps has been investigated
in a context of, e.g., topological superconductors in $(3+1)$
dimensions~\cite{Qi:2012cs,Stone:2016pof,Stalhammar:2021tcq}.
One of the characteristic features of this gapped phase 
is that there can be topological solitons, i.e., 
topologically stable objects, 
in addition to magnetic monopoles and axionic strings that always 
can exist.
When the photon is massive, there can be 
a quantized magnetic fluxes or vortex strings, which are 
called Abrikosov-Nielsen-Olesen 
(ANO) 
vortex strings~\cite{Abrikosov:1956sx,Nielsen:1973cs}.
When the axion is massive, there can be axionic domain walls, which connect distinct vacua of 
the axion, 
and a single axionic string is attached by 
some axionic domain walls~\cite{Sikivie:1982qv,Vilenkin:1982ks}.
The electromagnetic properties of the topological objects
have been investigated since they have been proposed.
For systems with massive photons, an ANO vortex string exhibits
the Aharonov-Bohm (AB) effect with a fractionally quantized phase
due to its quantized magnetic flux.
For systems with massive axions, 
an axionic domain wall has an induced electric charge 
when a magnetic flux is penetrated to the wall,
which we will call the Sikivie effect in this paper~\cite{Sikivie:1984yz}
(see also Refs.~\cite{Wilczek:1987mv,Qi:2008ew,Teo:2010zb}). 
Furthermore, the axionic domain wall exhibits an anomalous Hall effect:
When an electric flux, instead of a magnetic flux, 
is applied along an axionic domain wall,
the domain wall
has an induced electric current whose direction is perpendicular 
to the electric field~\cite{Sikivie:1984yz, Wilczek:1987mv, Qi:2008ew, Teo:2010zb}.

A natural question that arises is the following:
What is the underlying structure of these electromagnetic effects for the topological solitons?
The notion of extended symmetries may be one key ingredient to understand them.
Recently, symmetries for extended objects and topological solitons have been investigated in the language of higher-form symmetries;
higher $p$-form symmetries are symmetries under actions 
on $p$-dimensional extended objects~~\cite{Banks:2010zn, Kapustin:2014gua, Gaiotto:2014kfa} 
(see also Refs.~\cite{Batista:2004sc, Pantev:2005zs, Pantev:2005wj, Pantev:2005rh, Nussinov:2006iva, Nussinov:2008aa, Nussinov:2009zz, Nussinov:2011mz, Distler:2010zg}).
The conventional symmetries can be understood as 0-form symmetries,
since they act on local 0-dimensional operators.
In contrast, the AB effect in the gapped phase 
can be understood as a 2-form symmetry, 
where the charged object is a 
worldsheet
of a vortex line.
The higher-form symmetries have been applied to 
various systems in quantum field theories~%
\cite{Nussinov:2006iva,Nussinov:2009zz,Kapustin:2013uxa,Gaiotto:2017yup,Gaiotto:2017tne,Tanizaki:2017qhf,Tanizaki:2017mtm,Komargodski:2017dmc,Hirono:2018fjr,Hirono:2019oup,Hidaka:2019jtv,Anber:2019nze,Misumi:2019dwq,Anber:2020xfk,Anber:2020gig,Furusawa:2020kro}.

As in the conventional symmetries, higher-form symmetries can be
correlated to each other.
Their correlations can be elegantly described by $n$-groups~\cite{Sharpe:2015mja}.
Roughly speaking, an $n$-group is a set of groups 
for $0$-, ..., $(n-1)$-form symmetries with actions among them.
Quantum field theories with global 2- and 3-group symmetries 
have been investigated in Refs.~\cite{Kapustin:2013uxa,Kapustin:2013qsa,Bhardwaj:2016clt,Kapustin:2017jrc,Tachikawa:2017gyf, deAlmeida:2017dhy, Benini:2018reh,Cordova:2018cvg, Delcamp:2018wlb, Wen:2018zux, Delcamp:2019fdp, Thorngren:2020aph,Cordova:2020tij,Hsin:2019fhf,Hsin:2020nts,Gukov:2020btk,Iqbal:2020lrt,Brauner:2020rtz,DeWolfe:2020uzb,Brennan:2020ehu,Heidenreich:2020pkc,Apruzzi:2021vcu,Bhardwaj:2021wif}.
The higher-group symmetries can be efficiently found by 
coupling symmetry generators to background gauge fields 
for the higher-form symmetries.
In general, background gauging of a $p$-form symmetry
can be given by a $(p+1)$-form gauge field~\cite{Gaiotto:2014kfa}.
The $n$-group symmetry can be found by nontrivial 
gauge transformation laws between 1-, 2-,..., $n$-form gauge fields.

In particular, the $(3+1)$-dimensional axion electrodynamics with the massless axion and photon was found to be one of the simplest examples possessing
a 3-group structure~\cite{Hidaka:2020iaz,Hidaka:2020izy}.
In this system, there are 0-, 1-, 2-form symmetries 
associated to equations of motion and Bianchi identities for 
the axion and photon.
The 3-group structure has been found by correlation functions
between symmetry generators as well as a background gauging 
of the higher-form symmetries.
If the axion and photon become massive, we expect that 
the higher-from symmetries are different from those of the massless
axion and photon, since there can be symmetries associated to 
conservations of the ANO vortex strings and axionic domain wall 
in the gapped phase.
Thus, it is a nontrivial question what are higher-form symmetries
and associated higher-group symmetry in the axion electrodynamics in 
the gapped phase.

In the previous paper of the present authors~\cite{Hidaka:2021mml}, 
the low-energy effective action of the $(3+1)$-dimensional 
axion electrodynamics 
in the gapped phase was constructed.
Since the action only contains topological terms 
that do not depend on the metric of the spacetime, 
the effective theory is referred to as 
``topological axion electrodynamics.''
In this effective theory, 
it was found that there are 0-, 1-, 2-, and 3-form symmetries.
Furthermore, symmetry generators of the 0- and 1-form symmetries 
have nontrivial correlations,
similar to the current algebra of conventional symmetries.
A 2-form symmetry generator is induced on an intersection 
of the 0- and 1-form symmetry generators. 
Similarly, a 3-form symmetry generator is obtained on intersections of two 1-form symmetry generators.
Such nontrivial correlations are signals for higher-group symmetries, 
and in fact, a 4-group structure was found.

In this paper, we investigate higher-form symmetries in the topological
axion electrodynamics in more detail.
In particular, we discuss couplings of background gauge fields for the
higher-form symmetries.
We find that the gauging of each symmetry should be correlated.
The 1-form symmetry cannot be solely gauged with preserving the large
gauge invariance, and we need to gauge and modify the 3-form symmetry
simultaneously.
Further, the simultaneous gauging of the 0- and 1-form symmetries requires additional gauging and a modification of the 2-form symmetry.
We determine the modifications by the large gauge invariance.
By these modifications, we obtain a gauge theory of 1-, 2-, 3-, and 4-form gauge fields with correlations among them.

We then determine the higher-group structure 
of the topological axion electrodynamics by 
the modified background gauge fields.
The corresponding group is identified as a semi-strict 4-group or 3-crossed module~\cite{Arvasi:2009}.
We specify ingredients of the semi-strict 4-group by using the structure of the modified gauge fields.

By the background gauging, we show 't Hooft anomalies for the higher-form symmetries.
In general, the 't Hooft anomalies are obstructions to gauging global symmetries
dynamically~\cite{tHooft:1979rat,Frishman:1980dq,Coleman:1982yg}.
The presence of the 't Hooft anomalies forbids a symmetry-preserving gapped vacuum,
since such a vacuum does not have degrees of freedom that can match the anomalies.
The 't Hooft anomalies can be expressed as the ambiguity of the choice of a
5-dimensional space on which an action of the background gauge fields is
defined.
It has been known in many cases that 't Hooft anomalies in $D$ spacetime
dimensions can be canceled by adding a boundary of an appropriate
$(D+1)$-dimensional action given by
background gauge fields.
In our case, there are mixed 't Hooft anomalies between the pair of the
0- and 3-form symmetries as well as the pair of the 1- and 2-form
symmetries.
Furthermore, there is a mixed 't Hooft anomalies between the 0- and
1-form symmetries.
This type of the 't Hooft anomaly is called a 2-group anomaly, since it
depends on 0- and 1-form symmetries~\cite{Benini:2018reh}.
The 't Hooft anomalies can be expressed as a 5-dimensional action with
the background gauge fields of the higher-form symmetries.
We also discuss the physical consequences of the 't Hooft anomalies
such as topological order in the bulk and on the axionic domain wall.
While the essence of the physical effects has been discussed in 
the previous paper~\cite{Hidaka:2021mml},
we give detailed derivations of correlation functions with 
the intersections of symmetry generators.

This paper is organized as follows.
In section \ref{TAE}, we review the topological axion electrodynamics 
with a detailed derivation of the effective action.
The higher-form symmetries in the topological axion electrodynamics 
is then reviewed in section \ref{TAEHF}.
In section \ref{BG}, we consider the background gauging of 
the higher-form symmetries and 't Hooft anomalies.
In section \ref{BGphys}, we discuss physical consequences 
which can be derived by the background gauging.
Finally, we summarize this paper in section \ref{sum}.
In appendix \ref{corr}, we give detailed derivations of correlation functions for the symmetry generators discussed in section \ref{TAEHF}.

\section{\label{TAE}Topological axion electrodynamics}
In this section, we give an action of the axion electrodynamics 
where both the axion and photon are massive.
In the presence of the non-zero mass terms,
this theory is fully gapped.
The low-energy effective theory can be described by 
a topological field theory where 
the axion and photon are topologically 
coupled with 3- and 2-form gauge fields, respectively.
The topological field theory can be obtained by 
dual transformations.

\subsection{Action}
First, we introduce the action of the axion electrodynamics
with mass terms of the axion and photon. 
We begin with the following effective action
in a $(3+1)$-dimensional spacetime $M_4$,
\begin{equation}
    S
= 
- \int_{M_4} \(\fr{v^2}{2}|d\phi|^2 + V(\phi)\star 1  
+ \fr{1}{2e^2} |da|^2 + \fr{v'^2}{2}|d \chi - q a |^2 
-\fr{N}{8\pi^2} \phi da \wed da \).
\label{210424.1623}
\end{equation}
Here, we use the notation of differential forms. 
We introduce
$|\omega_n|^2  \coloneqq \omega_n \wed \star \omega_n
 = \omega_{\mu_1...\mu_n} \omega^{\mu_1...\mu_n} d^4x $ for 
a $n$-form field $\omega_n = ({1}/{n!})\, \omega_{\mu_1...\mu_n} dx^{\mu_1} \weds dx^{\mu_n} $, 
$ d $ is the exterior derivative, 
and $ \star $ denotes the Hodge star operator.
The quantities $v$ and $v'$ are mass dimension 1 parameters, 
the coupling constant $e$ is a dimensionless parameter,
and $N$ is an integer.
The axion $\phi$ is assumed to be a $2\pi $ 
periodic pseudo-scalar field,
\begin{equation}
 \phi ({\cal P})+ 2\pi \sim \phi ({\cal P}),
\end{equation}
for a point ${\cal P}$ in the spacetime $M_4$.
The periodicity can be regarded as gauge redundancy of the axion, 
i.e., the $2\pi$ shift of the
axion $\phi \to \phi + 2\pi$ is a gauge transformation. 
This redundancy is called a $(-1)$-form gauge symmetry~\cite{Kapustin:2014gua,Cordova:2019jnf,Cordova:2019uob}.
A gauge invariant object given by $\phi$ is a local point operator,
\begin{equation}
 L (q_\phi, {\cal P}) \coloneqq e^{iq_\phi \phi ({\cal P})}.
\end{equation}
Here, the charge of $q_\phi$ is quantized as 
\begin{equation}
 q_\phi \in \bb{Z}
\label{210702.1745}
\end{equation}
due to the $(-1)$-form gauge invariance.
Note that the axion operator $\phi ({\cal P})$ itself is 
not a gauge-invariant operator.
Since the axion is $2\pi$ periodic, the axion can have a nontrivial 
winding number along 
a 1-dimensional closed subspace ${\cal C}$:
\begin{equation}
 \int_{\cal C} d\phi \in 2\pi \bb{Z}.
\label{201205.1844}
\end{equation}
In other words, $\phi$ can be a multi-valued function.
Let us look at this quantization from a different angle.
We consider the two-point object of the 
axion $e^{i \phi ({\cal P})} e^{-i \phi ({\cal P}')}$
with the lowest charge $q_\phi = 1$, 
and express it by using a line integral along a line 
${\cal C_{P,P'}}$:
\begin{equation}
 e^{i \phi ({\cal P})} e^{-i \phi ({\cal P}')}
= e^{ i \int_{\cal C_{P,P'}} d \phi }.
\end{equation}
Here, the boundary of the line ${\cal C_{P,P'}}$ is 
${\cal P}$ and ${\cal P}'$: 
$\der {\cal C_{P,P'}} = {\cal P} \cup \b{\cal P}'$,
where $\b{\cal P}'$ is a point ${\cal P}' $ with the opposite 
orientation.
We have chosen the line ${\cal C_{P,P'}}$, 
but it is possible to choose another line ${\cal C'_{P,P'}} $
with the same boundaries
$\der {\cal C'_{P,P'}}  ={\cal P} \cup \b{\cal P}'$
as 
$ e^{i \phi ({\cal P})} e^{-i \phi ({\cal P})}
= e^{i\int_{\cal C'_{P,P'}} d \phi }$.
Since the two expressions should be identical, we have
the condition 
\begin{equation}
e^{ i \int_{\cal C_{P,P'}} d \phi } 
e^{- i\int_{\cal C'_{P,P'}} d \phi}
 = 
e^{ i \int_{\cal C} d \phi }  =1,
\end{equation}
where ${\cal C} = {\cal C_{P,P'}} \cup \b{\cal C}'_{\cal P,P'}$
is a loop without boundaries.
By this condition, we have the quantization in \er{201205.1844}.

The mass of the axion is given by the potential term
$V_k(\phi)$.
Since we are interested in the axionic domain wall, 
we assume that the potential term has $k$ of distinct minima
at $\phi = {2\pi n}/{k}$ ($n \in \bb{Z}$ mod $k$),
that is, the potential satisfies local stability conditions
$V'_{k} ({2\pi n}/{k}) = 0$
and 
$V''_{k} ({2\pi n}/{k}) =  M^4 >0$.
In addition, we choose the minimum of 
the potential as $V_k({2\pi n}/{k}) = 0$.
We further assume that 
the potential has a symmetry
under the shift 
$\phi \to \phi + {2\pi}/{k}$:
\begin{equation}
 V_k \(\phi + \fr{2\pi}{k}\) = V(\phi).
\label{210702.0328}
\end{equation}
Since each of the $k$ minima is physically different,
we regard this discrete transformation in \er{210702.0328} 
as a global symmetry.
A typical example is a cosine-type potential,
$V(\phi) \propto 1-\cos k\phi $, but we do not specify the 
detail of the potential because 
we will consider the low-energy limit.

The photon is given by a $U(1)$ 1-form gauge field 
$a = a_\mu dx^\mu$,
whose gauge transformation law is given by
\begin{equation}
 a \to a + d\lambda.
\label{210702.1550}
\end{equation}
Here, $\lambda$ is a $U(1)$ 0-form gauge parameter: 
it is a $2\pi $ 
periodic parameter $\lambda ({\cal P}) + 2\pi \sim \lambda ({\cal P})$,
and it can have a winding number, 
\begin{equation}
 \int_{\cal C} d\lambda \in 2\pi \bb{Z}.
\label{201203.1756}
\end{equation}
A gauge invariant object made of $a$ is a Wilson loop,
\begin{equation}
 W(q_a , {\cal C}) = e^{i q_a \int_{\cal C} a}.
\end{equation}
The invariance by a gauge parameter with a non-zero winding number 
in \er{201203.1756} requires
the quantization of the charge,
\begin{equation}
 q_a \in \bb{Z}.
\label{210702.1746}
\end{equation}
We can derive the flux quantization for the photon as 
in the case of the winding number of the axion.
The Wilson loop can be expressed using 
2-dimensional surfaces ${\cal S_C}$ and ${\cal S_C'}$ by the Stokes theorem as
\begin{equation}
 W(q_a , {\cal C}) 
 = e^{i q_a \int_{\cal S_C} da}
  = e^{i q_a \int_{\cal S_C'} da}.
\end{equation}
By the expressions, we have 
\begin{equation}
 e^{i q_a \int_{\cal S_C} da
 - iq_a\int_{\cal S_C'} da}
=
e^{i q_a \int_{\cal S} da}
  = 1,   
\end{equation}
 where ${\cal S} = {\cal S_C} \cup \b{\cal S}_{\cal C}'$
 is a closed 2-dimensional space, 
 and $\b{\cal S}_{\cal C}'$ is ${\cal S}_{\cal C}'$
 with an opposite orientation.
 The condition implies
 \begin{equation}
     \int_{\cal S} da \in 2\pi \bb{Z},
\label{201205.1646}
 \end{equation}
which means that there can be a magnetic monopole 
with a quantized charge.
This is the flux quantization condition for $a$.

The mass of the 
photon is given by the St\"uckelberg mechanism, 
which is a low-energy description of the Higgs mechanism 
without a radial mode.
This mechanism can be described by the scalar field $\chi$
with the charge $q \in \bb{Z}$,
which can be understood as a phase component of 
a charge $q$ Higgs field.
The gauge transformation law of $\chi $ 
under \er{210702.1550} is given by 
\begin{equation}
    \chi \to \chi + q \lambda, 
    \quad a \to a + d\lambda.
\end{equation}
By this gauge transformation, $\chi$ can be eaten 
by the gauge field $a$, and the gauge field becomes massive.

\subsection{Dual 2-form gauge theory}

Here, we dualize the action in \er{210424.1623} to the topological
action in the low-energy limit.
While the result has been shown in Ref.~\cite{Hidaka:2021mml},
we here discuss the dual transformation in detail.
In the energy scale lower than the masses of the axion and photon, 
there is no local excitation, but there can be topological excitation
such as an AB effect around the quantized magnetic vortices.
In order to see the topological effects, it will be convenient to 
dualize the theory to a topological field theory.
The topological field theory can be expressed by 
topological actions that consist of higher-form gauge fields.
In the absence of the axion, 
the low-energy effective theory around the 
ground state
can be described 
by a $BF$-theory given by 
1- and 2-form gauge fields~\cite{Horowitz:1989ng, Blau:1989bq}.
In the absence of the photon, we can describe the topological theory 
for the axion by 0- and 3-form gauge fields~\cite{Kapustin:2014gua,Gaiotto:2014kfa,Hidaka:2019mfm}.

First, we dualize the St\"uckelberg coupling to a $BF$-coupling.
We begin the following action that is written by 
the first-order derivative of $\chi$,
\begin{equation}
\begin{split}
    S'
&= 
- \int_{M_4} \Big(\fr{v^2}{2}|d\phi|^2 + V(\phi) \star1
+ \fr{1}{2e^2} |da|^2 + \fr{v'^2}{2}|w|^2 
-\fr{N}{8\pi^2} \phi da \wed da 
\\
&
\hph{\quad
- \int_{M_4} \Big(
}
+\fr{1}{2\pi} h \wed (w - d \chi +  q a ) \Big).
\end{split}
\end{equation}
Here, we have introduced 3- and 1-form fields 
$h$ and $w$, respectively.
The action is classically equivalent to the original 
action in \er{210424.1623}:
By the equation of motion for $h$, i.e., $w - d \chi + q a =0$,
the variables $w$ and $h$ can be eliminated, and 
we have the original action in \er{210424.1623}.
Instead, we can go to a dual theory by eliminating
 the scalar field $\chi$ 
and the 1-form field $w$ by their equations of motion.
The equation of motion for $\chi$ is $dh =0$, 
which can be locally solved by using a 2-form gauge field $b$ as
\begin{equation}
 h =  db.
\end{equation}
The 2-form gauge field has a gauge redundancy,
\begin{equation}
 b \to b + d\lambda_1,
\end{equation}
where $\lambda_1$ is a 1-form gauge parameter.
The normalization of the 2-form gauge field $b$ and 
$\lambda_1$ is given by
the quantization conditions,
\begin{equation}
 \int_{\cal V} db \in 2\pi \bb{Z}, 
\quad
 \int_{\cal S} d\lambda_1 \in 2\pi \bb{Z}. 
\label{210703.1429}
\end{equation}
Next, we eliminate the variable $w$. 
The equation of motion for $w$ is 
\begin{equation}
 v'^2 \star w - \fr{1}{2\pi} db =0 .
\end{equation}
Therefore, we have 
\begin{equation}
\begin{split}
    S_{BF}
&= 
- \int \Big(\fr{v^2}{2}|d\phi|^2 + V(\phi) \star 1
+ \fr{1}{2e^2} |da|^2 + \fr{1}{8\pi^2 v'^2}|db|^2 
-\fr{N}{8\pi^2} \phi da \wed da 
\\
&
\hph{\quad
- \int \Big(
}
- \fr{q}{2\pi} b \wed da  \Big).
\end{split}
\label{210627.1418}
\end{equation}
Thus, the scalar field $\chi$ is dualized to the 
2-form gauge field $b$.

\subsection{Dual 3-form gauge theory}

Next, we will dualize the potential term of the axion 
to a topological term given by the axion and a 3-form gauge field.
When we dualize the action in \er{210627.1418}, 
it will be convenient to include
the configuration of the domain wall,
since we can determine normalizations of dynamical fields
according to the configuration of the domain wall. 
In the low-energy region where we can neglect the width of the domain wall,
 we can express the configuration 
of the domain walls as the delta function 1-form,
\begin{equation}
    d\phi_{\rm W} = \fr{2\pi}{k} \delta_1 ({\cal V}),
\end{equation}
where ${\cal V}$ denotes the worldvolume of a domain wall.
Here, the delta function $(4-n)$-form on a $n$-dimensional subspace 
$\Sigma_{n}$ is defined by the relation,
\begin{equation}
 \int_{\Sigma_n} \omega_n = \int \omega_n \wed \delta_{4-n} (\Sigma_n).
\label{210809.1659}
\end{equation} 

Before dualizing the action,
 we decompose the axion into the fluctuation part
$\phi_{\rm F} $ and domain wall part $\phi_{\rm W}$,
\begin{equation}
    \phi = \phi_{\rm F} + \phi_{\rm W}.
\end{equation}
We can expand the potential term around 
$\phi_{\rm W}$.
Except for the place of the domain walls, 
we can set $V_k({2 \pi n}/{k}) = 0$.
By the local stability conditions,
$V_k'({2\pi n}/{k}) = 0$ and 
$V_k''({2\pi n}/{k}) = M^4$, 
the expansion of the potential up to the second order of $\phi_{\rm F}$ is
as follows:
\begin{equation}
    V_k(\phi)
     = \fr{1}{2} V''(\phi_{\rm W} ) \phi_{\rm F}^2 + {\cal O}(\phi_{\rm F}^3)
 = \fr{M^4}{2} (\phi - \phi_{\rm W})^2
 +{\cal O}(\phi_{\rm F}^3).
\end{equation}
Therefore, the action in \er{210627.1418} can be effectively 
written as 
\begin{equation}
\begin{split}
 S_{BF, {\rm quad}}
&= 
- \int_{M_4} \Big(\fr{v^2}{2}|d\phi|^2 + 
\fr{M^4}{2} (\phi - \phi_{\rm W})^2 \star 1
\\
&
\hph{\quad
- \int_{M_4} \Big(
}
+ \fr{1}{2e^2} |da|^2 + \fr{1}{8\pi^2 v'^2}|db|^2 
-\fr{N}{8\pi^2} \phi da \wed da 
- \fr{q}{2\pi} b \wed da ) \Big). 
\end{split}
\label{201204.2022}     
\end{equation}

Now, we dualize the action in \er{201204.2022}.
We replace $\phi_{\rm W} $ with a $2\pi$ periodic 
pseudo-scalar field $f$ by using a Lagrange multiplier 
3-form field $c$,
\begin{equation}
\begin{split}
    S_{BF, {\rm quad}}'
&     = 
-\int_{M_4} 
\Bigl(\fr{v^2}{2} |d\phi |^2 + \fr{M^4}{2}(\phi - f)^2 \star1
+ \fr{1}{2e^2} |da|^2 
-\fr{N}{8\pi^2} \phi da \wed da\\
&\quad+\frac{1}{8\pi^2v'^2}|db|^2-\frac{q}{2\pi}b\wedge da
- \fr{k}{2\pi}c \wed d (f- \phi_{\rm W} )
\Big).
\end{split}
\label{201204.2024}     
\end{equation}
Here, we assume that $f$ has the same boundary conditions as $\phi_{\rm W}$, $f|_{\rm bd} = \phi_{\rm W}|_{\rm bd}$, where the symbol 
`$|_{\rm bd}$' denotes the value at the boundary of the 
spacetime.
The 3-form field $c$ in the action in \er{201204.2024} 
can be regarded as
a $U(1)$ 3-form gauge field, 
since the action has the invariance under the gauge transformation,
\begin{equation}
    c \to c + d\lambda_2.
\end{equation}
Here, $\lambda_2$ is a $U(1)$ 2-form gauge parameter 
with the normalization,
\begin{equation}
    \int_{\cal V} d\lambda_2 \in 2\pi \bb{Z}.
    \label{201204.2048}
\end{equation}
The 3-form gauge field is also normalized on a closed 4-dimensional 
space $\Omega$ as 
\begin{equation}
    \int_{\Omega} dc \in 2\pi \bb{Z}.
\label{210703.1430}
    \end{equation}
The normalization of the Lagrange multiplier part $\fr{k}{2\pi}\int_{M_4} c \wed d (f-\phi_{\rm W})$
is determined so that it is invariant modulo $2\pi$
under the large gauge transformation in \er{201204.2048}.

We can go back to the original action in \er{201204.2022} 
by eliminating the 3-form gauge field $c$ 
using the equation of motion of $c$ with 
the boundary conditions of $f$ and $\phi_{\rm W}$.
Instead, we can go to the dual theory by eliminating
$f$ by its equation of motion,
\begin{equation}
     0= \fr{k}{2\pi} d c - M^4 (f- \phi)\star1.
\end{equation}
Substituting the equation into the action in \er{201204.2024},
we obtain the dual action,
\begin{equation}
\begin{split}
    S_{\rm dual, wall}
     &
     = 
     -\int_{M_4} 
\(\fr{v^2}{2} |d\phi |^2 
+ \fr{1}{2e^2} |da|^2 
+\frac{1}{8\pi^2v'^2}|db|^2
+
\fr{k^2}{8\pi^2 M^4} |dc|^2\)
\\
&\quad
+ \int_{ M_4}
\(\fr{k}{2\pi} c \wed d\phi 
+\frac{q}{2\pi}b\wedge da
+\fr{N}{8\pi^2} \phi da \wed da\)
- \int_{\cal V} c
\\
&\quad
+
\fr{k^2}{8\pi^2 M^4}\int_{M_4}
d (c \wed \star dc ).
    \end{split}
    \end{equation}
In the dual action, the potential term of the axion has been dualized to the quadratic kinetic term for the 3-form gauge field $|dc|^2$.%
\footnote{In our discussion, we have dualized the potential term
after expanding the potential term around the vacua.
It is possible to dualize the potential term without the expansion.
In this case, the detail of 
the potential is dualized to higher-derivative 
corrections to the kinetic term of the 3-form gauge field~\cite{Dvali:2005an,Nitta:2018yzb,Nitta:2018vyc}.}
Further, we obtain the topological term 
$c \wed d\phi$ between the axion and the 3-form gauge field.
Moreover, the worldvolume of the domain walls $d\phi_{\rm W} = ({2\pi}/{k}) \delta_1 ({\cal V})$
is now electrically coupled with the 
3-form gauge field as $\int_{\cal V} c$.
The normalization of the 3-form gauge field $c$ is determined so
that a single domain wall has a unit charge of the 3-form gauge field.
The last term is the boundary term for the kinetic term, which is generally needed to 
have an energy-momentum tensor consistent with 
the equation of motion~\cite{Brown:1987dd, Brown:1988kg, Duff:1989ah, Duncan:1989ug}.

In a sufficiently lower energy scale than the masses of the axion and photon, we can neglect the kinetic terms of $\phi$, $a$, $b$, and $c$.
We thus arrive at the following topological action,
\begin{equation}
    S_{\rm TAE}
= 
\int \( \fr{k}{2\pi} c \wed  d\phi 
+ \fr{q}{2\pi} b \wed da
+
\fr{N}{8\pi^2} \phi da \wed da\) .
\label{210514.1613}
\end{equation} 
Following the previous paper of the present authors~\cite{Hidaka:2021mml}, 
we call this theory the ``topological axion electrodynamics,''
since the action does not depend on the metric in the spacetime.

\section{\label{TAEHF}Higher-form symmetries in topological axion electrodynamics}
In this section, we review higher-form global symmetries in 
the topological axion electrodynamics~\cite{Hidaka:2021mml}.
The higher-form symmetries are found by the equations of motion and Bianchi identities of the dynamical fields.

\subsection{\label{elec}Electric symmetries}

First, we show higher-form symmetries associated with the equations 
of motion.
Following Ref.~\cite{Gaiotto:2014kfa}, 
we will call them electric symmetries,
but we will often omit ``electric'' if there is no confusion.
The equations of motion
 for the dynamical fields, i.e., $\phi$, $a$, $b$, and $c$ 
are
\begin{equation}
\begin{split}
&
 \fr{k}{2\pi} dc + \fr{N}{8\pi^2} 
 da \wed da  =0,
\quad
\fr{q}{2\pi}  db +\fr{N}{4\pi^2} d\phi \wed d a
 =0, 
\quad
\fr{q}{2\pi} da  = 0, 
\quad
\fr{k}{2\pi} d\phi =0,
\end{split}
\end{equation}
respectively.
The corresponding symmetry generators have the form,
\begin{align}
  U_0 (e^{i \alpha_0 },{\cal V})
 &= \exp \(- i \alpha_0
\int_{\cal V} \(\fr{k}{2\pi }c + \fr{N}{8\pi^2} a \wed da \)\),
\label{210702.1720-0}
\\
U_1(e^{i \alpha_1 }, {\cal S})
& = \exp \(
- i\alpha_1 \int_{\cal S}
 \(\fr{q}{2\pi } b + \fr{N}{4\pi^2}\phi da
\)
\),
\label{210702.1720-1}
\\
 U_2 (e^{i \alpha_2 },{\cal C})
&= \exp \(- i \alpha_2 \int_{\cal C} \fr{q}{2\pi } a\),
\label{210702.1720-2}
\\
U_3 (e^{i\alpha_3 },{\cal (P,P')})
&
= \exp \(- i\alpha_3 \cdot \fr{k}{2\pi} (\phi ({\cal P})-\phi({\cal P'}) ) \),
\label{210702.1720-3} 
\end{align}
where $e^{i\alpha_0},\cdots ,e^{i\alpha_3}$ may be 
$U(1)$ parameters, which will be determined below.
Hereafter, we assume that 
the subspaces ${\cal V}$, ${\cal S}$, and ${\cal C}$ on which 
the symmetry generators are defined 
do not have self-intersections for simplicity.

As we explained in \ers{210702.1745} and \eqref{210702.1746},
the parameters $e^{i\alpha_2}$ and $e^{i\alpha_3}$ are 
constrained as
\begin{equation}
 e^{i\alpha_2}  = e^{2\pi i n_2/q} \in \bb{Z}_q,
\quad
 e^{i\alpha_3}  = e^{2\pi i n_3/k} \in \bb{Z}_k,
\end{equation}
respectively.
Further, the parameters $e^{i\alpha_0}$ and $e^{i\alpha_1}$ 
are also subject to some constraints due to the large gauge invariance of 
the integrals.
To make the integrand gauge invariant, we define $U_0$ and $U_1$
by using the Stokes theorem,
\begin{equation}
\begin{split}
  U_0 (e^{i \alpha_0 },{\cal V})
 &= \exp \(- i \alpha_0
\int_{\Omega_{\cal V}} \(\fr{k}{2\pi }dc + \fr{N}{8\pi^2} da \wed da \)\),
\\
U_1(e^{i \alpha_1 }, {\cal S})
& = \exp \(
- i\alpha_1 \int_{\cal V_S}
 \(\fr{q}{2\pi } db + \fr{N}{4\pi^2}d\phi \wed da
\)
\).
\end{split}
\label{210702.2028}
\end{equation}
Here, $\Omega_{\cal V}$ and ${\cal V_S}$ are 4- and 3-dimensional 
manifolds whose boundaries are ${\cal V}$ and ${\cal S}$, 
respectively.
By the Stokes theorem, 
we have the manifestly gauge invariant integrands.
However, we have chosen auxiliary spaces by hand.
Therefore, we require that the symmetry generators should be independent of the choices of the auxiliary spaces.
To see the conditions that satisfy the requirement, 
we choose another subspaces $\Omega_{\cal V}' $ and 
${\cal V_S'}$ for the 0- and 1-form symmetry generators
which satisfy $\der \Omega_{\cal V}' = {\cal V}$ 
and $\der {\cal V_S'}= {\cal S}$, 
respectively.
The independence of the choices can be expressed by 
the following integrals on 
closed 4- and 3-dimensional spaces
$\Omega = \Omega_{\cal V} \cup \b\Omega_{\cal V}'$
and ${\cal V} = {\cal V_S} \cup \b{\cal V}_{\cal S}'$
as
\begin{equation}
\begin{split}
 \exp \(- i \alpha_0
\int_{\Omega} \(\fr{k}{2\pi }dc + \fr{N}{8\pi^2} da \wed da \)\)
&=1,
\\
 \exp \(
- i\alpha_1 \int_{\cal V}
 \(\fr{q}{2\pi } db + \fr{N}{4\pi^2}d\phi \wed da
\)
\)
&
=1
.
\end{split}
\label{210703.0254}
\end{equation}
Because of the flux quantization conditions
in 
\ers{201205.1844}, 
\eqref{201205.1646}, 
\eqref{210703.1429}, 
\eqref{210703.1430}, 
and 
$\int_{\Omega} da \wed da \in 2 \cdot (2\pi)^2 \bb{Z}$ 
on a spin manifold, 
we find that 
the parameters $e^{i \alpha_0 }$ and $e^{i \alpha_1 }$
should belong to discrete groups,
\begin{equation}
 e^{i \alpha_0 } = e^{2\pi i n_0/m} \in \bb{Z}_m,
\quad
 e^{i \alpha_1 } = e^{2\pi i n_1/p} \in \bb{Z}_p,
\end{equation}
where
we have defined $m = \gcd (N,k)$ and 
$p = \gcd(N,q)$. 
The symbol ``$\gcd$'' stands for the greatest common divisor.

To summarize, the gauge-invariant symmetry generators
are given by
\begin{align}
   U_0 (e^{2\pi i n_0/m},{\cal V})
 &= \exp \(- 2\pi i \fr{ n_0}{m} 
\int_{\cal V} \(\fr{k}{2\pi }c + \fr{N}{8\pi^2} a \wed da \)\),
\label{210702.1727-0}
\\
U_1(e^{2\pi i n_1/p}, {\cal S})
& = \exp \(
- 2\pi i \fr{ n_1}{p} \int_{\cal S}
 \(\fr{q}{2\pi } b + \fr{N}{4\pi^2}\phi da
\)
\),
\label{210702.1727-1}
\\
 U_2 (e^{2\pi i n_2/q},{\cal C})
&= \exp \(- in_2 \int_{\cal C} a\),
\label{210702.1727-2}
\\
U_3 (e^{2\pi i n_3/k},{\cal (P,P')})
&
= \exp \(- in_3 (\phi ({\cal P})-\phi({\cal P'}) ) \).
\label{210702.1727-3}
\end{align}
They form
$\bb{Z}_m$ 0-form, $\bb{Z}_p$ 1-form, $\bb{Z}_q$ 2-form, 
and $\bb{Z}_k$ 3-form global symmetries.

The charged objects on which the symmetry generators act 
are the Wilson loop 
and its analogues, given by
\begin{align}
 L (q_0, {\cal P})
 &=  e^{i q_0 \phi ({\cal P})},
\\
 W(q_1, {\cal C}) 
&
= e^{iq_1 \int_{\cal C} a} = 
U_2(e^{-2\pi i q_1/q},{\cal C}), 
\\
 V(q_2, {\cal S}) &
= e^{iq_2 \int_{\cal S} b}, 
\\
 D(q_3, {\cal V}) &
= e^{i q_3 \int_{\cal V} c},  
\end{align}
respectively.
Here, the charges are integers $q_0,..., q_3 \in \bb{Z}$ 
because of the large gauge invariance of charged objects.
We remark that $W (q_1, {\cal C})$ is identical to the symmetry generator
$U_2(e^{2\pi i q_1/q}, {\cal C})$.
We will use this property to show that the topological axion electrodynamics
is topologically ordered.
The symmetry transformations are found by the correlation functions
(see Appendix \ref{corr} for derivations),
\begin{align}
 \vevs{  U_0 (e^{2\pi i n_0/m},{\cal V}) L (q_0, {\cal P}) }
&= e^{2\pi i q_0 n_0 \link ({\cal V,P}) /m } \vevs{ L (q_0, {\cal P}) },
\label{210516.1529-0}
\\ 
 \vevs{  U_1 (e^{2\pi i n_1/p},{\cal S}) W (q_1, {\cal C}) }
&= e^{2\pi i q_1 n_1 \link ({\cal S,C}) /p } 
\vevs{ W (q_1, {\cal C}) },
\label{210516.1529-1}
\\ 
 \vevs{  U_2 (e^{2\pi i n_2/q},{\cal C}) V (q_2, {\cal S}) }
&= e^{2\pi i q_2 n_2 \link ({\cal S,C}) /q } 
\vevs{ V (q_2, {\cal S}) },
\label{210516.1529-2}
\\ 
 \vevs{  U_3 (e^{2\pi i n_3/q},({\cal P,P'})) D (q_3, {\cal V}) }
&= e^{2\pi i q_3 n_3 \link ({\cal (P,P'),V}) /k } 
\vevs{D (q_3, {\cal V}) }. 
\label{210516.1529-3}
\end{align}
Here, we have defined a linking number between 
$n$- and $(3-n)$-dimensional subspaces
$\Sigma_n$ and $\Sigma_{3-n}'$ as
\begin{equation}
\begin{split}
 \link (\Sigma_n, \Sigma'_{3-n}) 
= 
\int_{\Omega_{\Sigma_n}} \delta_{n+1} (\Sigma'_{3-n})
= 
\int_{M_4} 
\delta_{n+1} (\Sigma'_{3-n}) 
\wed 
\delta_{3-n} (\Omega_{\Sigma_n}), 
\end{split}
\end{equation} 
where $\Omega_{\Sigma_n}$ is an 
$(n+1)$-dimensional subspace whose boundary is $\Sigma_n$.
\subsection{Magnetic symmetries}
In addition, we have the following symmetry generators 
associated to the Bianchi identities for the dynamical fields,
\begin{equation}
\begin{split}
 U_{\rm 2M} (e^{i\beta_2 }, {\cal C}) 
&
= e^{i\beta_2 \int_{\cal C} \frac{d\phi}{2\pi}},
\quad
 U_{\rm 1M} (e^{i\beta_1 }, {\cal S})
 = e^{i\beta_1 \int_{\cal S} \frac{da}{2\pi}},
\\
 U_{\rm 0M} (e^{i\beta_0 }, {\cal V})
&
 =e^{i\beta_0\int_{\cal V} \frac{db}{2\pi}},
\quad
 U_{\rm -1M} (e^{i\beta_{-1} }, \Omega)
 = e^{i\beta_{-1} \int_{\Omega} \frac{dc}{2\pi}}.
 \end{split}
\end{equation}
Here, $e^{i\beta_2}$,..., $e^{i\beta_{-1}}$ are $U(1)$ parameters.
We will call these symmetries magnetic symmetries, since 
they are associated with the Bianchi identities.

The charged objects for the 2-, 1-, 0- form symmetries 
are
a worldsheet of the axionic string with the winding number 
$q_{\rm 2M}$ denoted as $S (q_{\rm 2M},{\cal S})$,
a charge $q_{\rm 1M}$ 't Hooft loop
$T(q_{\rm 1M},{\cal C})$,
a pair of charge $\pm q_{\rm 0M} $ instantons
 $I (q_{\rm 0M}, ({\cal P, P'}))$,
respectively.
We should remark that 
the charged objects for the 
magnetic symmetries should
be boundaries of the electric symmetry 
generators, 
but we do not write the configurations 
of the electric symmetry generators since the configurations 
of the magnetic objects do 
not depend on them.
Note that we do not consider a magnetic object for the 3-form gauge field,
since the spacetime dimension of the object would be $-1$.
The symmetry transformation laws are 
\begin{equation}
\begin{split}
 \vevs{  U_{\rm 0M} (e^{i\beta_0 } ,{\cal V}) I (q_{\rm 0M}, ({\cal P, P'})) }
&= e^{i\beta_0  q_{\rm 0M} \link ({\cal V,(P, P')}) } 
\vevs{ I (q_{\rm 0M}, ({\cal P, P'})) },
\\ 
 \vevs{  U_{\rm 1M} (e^{i\beta_1},{\cal S}) T (q_{\rm 1M}, {\cal C}) }
&= e^{i q_{\rm 1M} \beta_1 \link ({\cal S,C}) } 
\vevs{ T (q_{\rm 1M}, {\cal C}) },
\\ 
 \vevs{  U_{\rm 2M} (e^{i\beta_2},{\cal C}) S (q_{\rm 2M}, {\cal S}) }
&= e^{ i q_{\rm 2M } \beta_2  \link ({\cal S,C}) } 
\vevs{ S (q_{\rm 2M}, {\cal S}) }.
\end{split}
\end{equation}
In addition, there are $U(1)$ 0- and $(-1)$-form symmetries
 given by products of the currents for the magnetic symmetries,
\begin{align}
 U_{0 {\rm CW} } (e^{i\gamma_{0} }, {\cal V}) 
&
= \exp\(i\gamma_{0} \int_{\cal V} \fr{d\phi}{2\pi} \wed \fr{da}{2\pi}\),
\\
 U^1_{-1 {\rm CW} } (e^{i\gamma_{-1} },\Omega) 
&
= \exp\(i\gamma^1_{-1} \int_{\Omega} 
\fr{1}{2} \fr{da}{2\pi} \wed \fr{da}{2\pi}\),
\\
 U^2_{-1 {\rm CW} } (e^{i\gamma_{-1} },\Omega) 
&
= \exp\(i\gamma^2_{-1} \int_{\Omega} 
 \fr{d\phi }{2\pi} \wed \fr{db}{2\pi}\).
\end{align}
They are called Chern-Weil global symmetries~\cite{Heidenreich:2020pkc,Brauner:2020rtz}.

\section{\label{BG}Background gauging and 't Hooft anomalies}

In this section, we discuss the background gauging 
of the higher-form global symmetries discussed in 
the previous section.
The correlations between the symmetry generators 
can be efficiently discussed by the background gauging.
By the background gauging, we show that 
the higher-form symmetries of the topological axion electrodynamics 
possesses a semi-strict 4-group structure.
Furthermore, we can find possible 't Hooft anomalies for the 
higher-form global symmetries, which are obstructions to 
gauge the symmetries dynamically.

\subsection{Modification of background gauging}

We consider the background gauging of the higher-form symmetries
by introducing appropriate background gauge fields.
Before performing the background gauging, it is useful to rewrite the action~\eqref{210514.1613} by one defined on the boundary of an auxiliary 5-dimensional manifold $X_5$:
\begin{equation}
    S_{\rm TAE} [X_5]
= 
\int_{X_5} \( \fr{k}{2\pi} d c \wed  d\phi 
+ \fr{q}{2\pi} d b \wed da
+
\fr{N}{8\pi^2} d\phi \wed da \wed da\) \quad {\rm mod} \, 2\pi,
\end{equation} 
with $\der X_5 =M_4$. 
Hereafter, we omit ``mod $2\pi$'' 
of the actions given by 5-dimensional manifolds
which does not contribute to $e^{ iS_{\rm TAE} [X_5]}$ 
in the path integral.
This action is manifestly gauge invariant,
reducing to the original one in Eq.~\eqref{210514.1613} with the help of the Stokes theorem.
Furthermore, the action does not depend on the choice of the 5-dimensional manifold $X_5$.
To show the independence, we choose another 5-dimensional manifold
$X_5'$ satisfying $\der X_5' = M_4$. 
The difference between these two choices in the path integral 
can be evaluated as
\begin{equation}
     e^{i S_{\rm TAE} [X_5]} e^{-i S_{\rm TAE} [X_5']}
 = e^{ i S_{\rm TAE} [Z_5]}  = 1,
\end{equation}
where $Z_5 = X_5 \cup \b{X}_5'$ is a 5-dimensional manifold
without boundaries $\der Z_5  =\emptyset$.
Therefore, the action $ S_{\rm TAE} [X_5]$ 
does not depend on the choice of $X_5$ mod $2\pi$.

Now, we couple the action $S_{\rm TAE} [X_5]$ to background gauge fields.
We first consider the electric symmetries discussed in section \ref{elec}.
Since these higher-form symmetries correspond to shift symmetries of dynamical fields,
the background gauge fields can be coupled with the dynamical fields by St\"uckelberg couplings.

For example, for the $\mathbb{Z}_m$ 0-form symmetry, we may replace
$d\phi$ by $d\phi-A_1^m$ with a $\mathbb{Z}_m$ 1-form gauge field
$A_1^m$.
Here, the $\mathbb{Z}_m$ 1-form gauge field means that $A_1^m$ is
closed, $dA_1^m =0$, and normalized as 
$\int_{\mathcal{C}} A^m_1 \in \frac{2\pi}{m}\mathbb{Z}$ 
on a one-dimensional closed path
$\mathcal{C}$.
In other words, $A_1^m$ can be locally expressed as 
\begin{equation}
    m A^m_1=d A_0^m
    \label{eq:gauge field A0},
\end{equation}
where $A_0^m$ is a 0-form gauge field with the normalization $\int_{\mathcal{C}}dA_0^m\in 2\pi\mathbb{Z} $.
The combination $d\phi-A_1^m$ is gauge invariant under 
\begin{equation}
 \phi \to \phi + \Lambda_0,
\quad
A_1^m \to A_1^m + d\Lambda_0,
\quad
A_0^m \to A_0^m + m \Lambda_0,
\label{210515.0144}
\end{equation}
where $\Lambda_0$ is a gauge parameter satisfying $\int_{\cal C} d\Lambda_0\in 2\pi\mathbb{Z}$.
Similar background gauging can be performed for the other higher-form symmetries. Therefore, a naive gauging would be given by 
\begin{equation}
\begin{split}
    S_{\rm TAE, 0} [X_5]
&
= 
\int_{X_5} \Big( \fr{k}{2\pi}
\(dc - D^k_4\) 
\wed  (d\phi -A^m_1) 
+ \fr{q}{2\pi} (db - C^q_3) \wed (da - B^p_2)
\\
&
\quad
\hph{\int_{X_5} \Big( }
+
\fr{N}{8\pi^2} (d\phi - A^m_1) \wed (da - B^p_2) \wed (da -B^p_2)\Big) . 
\label{eq:naive action}
\end{split}
\end{equation} 
Here, we have introduced the gauge fields 
$B^p_2$, $C^q_3$, and 
$D^k_4$, which  are 
$\bb{Z}_p$ 2-form, 
$\bb{Z}_q$ 3-form,
and 
$\bb{Z}_k$ 4-form 
gauge fields satisfying
\begin{equation}
 p B^p_2 = dB^p_1,
\quad
 q C^q_3 = dC^q_2,
\quad
 k D^k_4 = dD^k_3,
\label{210704.1540}
\end{equation}
with 
1-, 2-, 3-form gauge fields,
$B^p_1$, $C^q_2$, and 
$D^k_3$, respectively.
The gauge transformation laws are given by
\begin{align}
 &
a \to a + \Lambda_1,
\quad
B_2^p \to B_2^p + d\Lambda_1,
\quad
B_1^p \to B_1^p + p \Lambda_1,
\label{210515.0145}
\\
&
 b \to b + \Lambda_2,
\quad
C_3^q \to C_3^q + d\Lambda_2,
\quad
C_2^q \to C_2^q + q \Lambda_2,
\label{210515.0146}
\\
& 
c \to c + \Lambda_3,
\quad
D_4^k \to D_4^k + d\Lambda_3,
\quad
D_3^k \to D_3^k +  k \Lambda_3.
\label{210515.0147}
\end{align}
Since $A_0^m$, $B_1^p$, $C_2^q$ and $D_3^k$ are also gauge fields, they
transform under their gauge transformations:
\begin{align}
A_0^m&\to A_0^m+2\pi,\\
B_1^p&\to B_1^p+d\Lambda^p_0,\\
C_2^q&\to C_2^q+d\Lambda^q_1,\\
D_3^k&\to D_3^k+d\Lambda^k_2.
\end{align}
These gauge fields are coupled to magnetic and Chern-Weil symmetries whose currents are 
\begin{align}
A_0^m&:\frac{k}{m}\frac{dc}{2\pi}+\frac{N}{m} \frac{1}{2}\frac{da}{2\pi}\wedge \frac{da}{2\pi},\\
B_1^p&:\frac{q}{ p}\frac{db}{2\pi} + \frac{N}{p}\frac{d\phi}{2\pi}\wedge \frac{da}{2\pi},\\
C_2^q&:\frac{da}{2\pi}, \label{eq:C2q coupling}\\
D_3^k&:\frac{d\phi}{2\pi}. \label{eq:D3k coupling}
\end{align}
We can see that $A_0^m$ and $B_2^p$ are coupled to linear combinations
of magnetic and Chern-Weil symmetries, while $C_2^q$ and $D_3^k$ are
directly coupled to magnetic ones.

The gauge fields and gauge parameters are normalized as
\begin{align}
 &\int_{\cal S} dB_1^p,
\,
 \int_{\cal V} dC_2^q,
\,
 \int_{\Omega} dD_3^k \in 2\pi \bb{Z},
\\
\label{210515.0211}
& \int_{\cal S} d\Lambda_1,
\,
 \int_{\cal V} d\Lambda_2,
\,
 \int_{\Omega} d\Lambda_3,
\int_{\cal C} d\Lambda_0^p,
\int_{\cal S} d\Lambda_1^q,
\int_{\cal V} d\Lambda_2^k,
 \in 2\pi \bb{Z},
\end{align}
respectively.
Similarly to the $\mathbb{Z}_m$ $0$-form symmetry, 
the background gauge fields 
are flat, but they 
can have fractional AB phases,
\begin{equation}
 \begin{split}
  \int_{\cal S} B_2^p \in \fr{2\pi}{p} \bb{Z},
\quad
  \int_{\cal V} C_3^q \in \fr{2\pi}{q} \bb{Z},
\quad
  \int_{\Omega} D_4^k \in \fr{2\pi}{k} \bb{Z}.
 \end{split}
\end{equation}

However, the naive action in Eq.~\eqref{eq:naive action} has ambiguity in the choice of the 
auxiliary manifold $X_5$
if $N/ (mp)$ or $N/p^2$ are nontrivial fractional numbers.
Such ambiguity can be found by evaluating the 
difference of the background actions between the two choices of 
5-dimensional manifolds $X_5$ and $X_5'$:
 \begin{equation}
\begin{split}
&
S_{\rm TAE,0} [X_5] - S_{\rm TAE, 0} [X_5']
 = 
S_{\rm TAE,0} [Z_5]
\\
&
=
\int_{Z_5} \Big( \fr{k}{2\pi}
\(dc - D^k_4\) 
\wed  (d\phi -A^m_1) 
+ \fr{q}{2\pi} (db - C^q_3) \wed (da - B^p_2)
\\
&
\quad
\hph{\int_{X_5} \Big( }
+
\fr{N}{8\pi^2} (d\phi - A^m_1) \wed (da - B^p_2) \wed (da -B^p_2)\Big) 
\\
&=
\int_{Z_5} \Big( \fr{k}{2\pi}
D^k_4
\wed A^m_1
+ \fr{q}{2\pi}C^q_3  \wed B^p_2
-
\fr{N}{8\pi^2}A^m_1 \wed  B^p_2  \wed B^p_2
\\
&
\quad
\hph{\int_{X_5} \Big( }
+
\fr{N}{8\pi^2} d\phi \wed B^p_2 \wed B^p_2 
+
\fr{N}{4\pi^2} A^m_1 \wed da \wed B^p_2 \Big).
 \end{split}
 \label{210818.1325}
\end{equation}
We have dropped terms proportional to $2\pi$ in the last equality, which do not contribute to the weight of the path integral $e^{i S_{\rm TAE,0}}$.
The first line in the last equation that is independent of dynamical fields
may represent possible 't Hooft anomalies.
However, the second line, depending on the 
dynamical fields $\phi$ and $a$,
leads to the inconsistency 
of the theory
if $N / p^2$ or $N/ (mp)$ are fractional. 
The inconsistency can also be understood as a 
violation of the large gauge invariance in 
the 4-dimensional spacetime~\cite{Benini:2018reh,Hidaka:2020izy}.
To preserve the consistency of the theory, 
we should modify the background fields such that 
there is no ambiguity of the choice of the 5-dimensional manifolds for terms containing dynamical fields in the action.
Since the ambiguity is caused by the terms proportional to 
$da$ and $d\phi $, we will modify the 3- and 4-form gauge fields.

We modify the 3-form gauge field $C^q_3$ and 
4-form gauge field $D_4^k$ as follows,
\begin{align}
 C_3^q 
&
\to C_3^{Q} \coloneqq C_3^q + \fr{N}{2\pi q} A_1^m \wed B_2^p,
 \label{210709.1701}
\\
 D_4^k 
&
\to D_4^{K}\coloneqq D_4^k + \fr{N}{4\pi k} B_2^p \wed B_2^p.
\label{210704.1558}
\end{align}
Or equivalently, we can write
\begin{align}
 dC_2^q +\fr{N}{2\pi } A_1^m \wed B_2^p 
= 
qC_3^Q,
\label{eq: modified gauge fields1}
\\
 dD_3^k +  \fr{N}{4\pi } B_2^p \wed B_2^p 
= 
kD_4^K.
\label{eq: modified gauge fields2}
\end{align}
As we will discuss in section~\ref{sec:higher group structure}, these
are the key equations in our higher-order group.

We should preserve the gauge invariance of the St\"uckelberg couplings
$db - C_3^{Q}$ and $dc - D_4^K$.  We thus impose the modified gauge
transformation laws
\begin{align}
\begin{split}
C_3^q  
&
\to 
C_3^q  +d\Lambda_2
- \fr{N}{2\pi q} (d\Lambda_0 \wed B_2^p + A_1^m \wed d\Lambda_1 
+ d\Lambda_0 \wed d\Lambda_1),
\\
C_2^q 
 &
\to C_2^q  +q\Lambda_2
- \fr{N}{2\pi} (\Lambda_0 B_2^p + \Lambda_1 \wed A_1^m  
+ \Lambda_0 d\Lambda_1),
\end{split}
\label{210703.2232}
\\
 \begin{split}
D_4^k 
&
\to D_4^k +d\Lambda_3
-  \fr{N}{2 \pi k} d\Lambda_1 \wed B^p_2
-  \fr{N}{4\pi k}d\Lambda_1 \wed d\Lambda_1,
\\
D_3^k
 &
\to D_3^k +k\Lambda_3
-  \fr{N}{2 \pi } \Lambda_1 \wed B^p_2
-  \fr{N}{4\pi }\Lambda_1 \wed d\Lambda_1.
 \end{split}
\label{210703.2233}
\end{align}
We remark that the modifications of the 
gauge transformation laws do not violate 
the $2\pi$ periodicity of $C_2^q$ and $D_3^k$ in \er{210515.0211}.
 
By the modifications of the background 3- and 4-form gauge fields,
their fractional AB phases are modified as
\begin{equation}
 \int_{\cal V} C_3^{Q} = \fr{2\pi}{q}
 \( n_2 + \fr{N}{mp} n_{01}
 \),
\quad 
 \int_{\Omega} D_4^{K} = \fr{2\pi }{k } \( n_3 + \fr{N}{p^2} n_{11}^2 \).
\end{equation}
Here, 
$n_2 = \fr{1}{2\pi}\int_{\cal V} dC^q_2$,
$n_{12} = \fr{1}{4\pi^2}\int_{\cal V} dA_0^m \wed dB_1^p$,
$n_{11} = \fr{1}{8\pi^2}\int_{\Omega} dB_1^p \wed dB_1^p$
are integers.
Thus, the gauge fields 
$ C_3^{Q}$ and $D_4^{K}$ are 
$\bb{Z}_{Q}$ 
and $\bb{Z}_{K}$ gauge fields, 
where the integers $Q$ and $K$ are 
defined by
\begin{equation}
 Q = q \cdot \fr{mp }{\gcd(N, mp )} ,
\quad
 K = k\cdot \fr{p^2}{\gcd(N, p^2)} ,
\end{equation}
respectively.
Here, the integers $mp / \gcd(N, mp )$ and 
$p^2 / \gcd(N, p^2)$ are denominators of 
$N/ (mp)$ and $N / p^2$, which characterize
the necessity of the modifications of the background gauge fields as in \er{210818.1325} if they are nontrivial.

After the above modifications, the gauged action becomes
\begin{equation}
\begin{split}
    S_{\rm TAE, bg}[X_5]
&
= 
\int_{X_5} \Big( \fr{k}{2\pi}
\(dc - D^K_4\) 
\wed  (d\phi -A^m_1) 
+ \fr{q}{2\pi} (db - C^Q_3) \wed (da - B^p_2)
\\
&
\quad
\hph{\int_{X_5} \Big( }
+
\fr{N}{8\pi^2} (d\phi - A^m_1) \wed (da - B^p_2) \wed (da -B^p_2)\Big) 
\\
&
= 
\int_{X_5} \Big( \fr{k}{2\pi}
\(dc - D^k_4\) 
\wed  (d\phi -A^m_1) 
+ \fr{q}{2\pi} (db - C^q_3) \wed (da - B^p_2)
\\
&
\quad
\hph{\int_{X_5} \Big( }
+
\fr{N}{8\pi^2} (d\phi - A^m_1) \wed da  \wed da
-
\fr{N}{4\pi^2} d\phi \wed da  \wed B^p_2
\\
&
\quad
\hph{\int_{X_5} \Big( }
+ \fr{N}{4\pi^2} A_1^m \wed B_2^p \wed B^p_2
\Big) . 
\end{split}
\label{210514.2032}
\end{equation} 
This action causes no ambiguity due to the dynamical fields.

We can further gauge the magnetic symmetries.
However, further gauging might cause redundancy because $A_0^m$,
$B_1^p$, $C_2^q$, and $D_3^k$ couple to magnetic symmetries.
In particular, $C_2^q$, and $D_3^k$ are directly coupled to magnetic
$1$- and $2$-form symmetries shown in Eqs.~\eqref{eq:C2q coupling} and
\eqref{eq:D3k coupling}.
Let us look at this redundancy in detail by gauging the magnetic
symmetries.
In the absence of the background gauge fields for the electric
symmetries, the background gauging of the magnetic symmetries are given
by adding the following action to $S_{\rm TAE}$,
\begin{equation}
 S_{\rm bg,M} [X_5]
= 
\fr{1}{2\pi}\int_{X_5}
 (dc \wed d\Phi_0^{\rm M}
-
 db \wed d A_1^{\rm M} 
+
 da \wed d B_2^{\rm M} 
-
d\phi  \wed d C_3^{\rm M}). 
\end{equation}
Here, $\Phi_0^{\rm M}$, $A_1^{\rm M}$, $B_2^{\rm M}$, and $C_3^{\rm M}$
are 0-, 1-, 2-, and 3-form $U(1)$ gauge fields.
The gauge transformations of the gauge fields are
\begin{equation}
 \Phi_0^{\rm M} \to \Phi_0^{\rm M} + 2\pi, 
\quad
A_1^{\rm M} \to A_1^{\rm M} + d \Lambda_0^{\rm M}, 
\quad
B_2^{\rm M} \to B_2^{\rm M} + d \Lambda_1^{\rm M}, 
\quad
C_3^{\rm M} \to C_3^{\rm M} + d \Lambda_2^{\rm M}.
\end{equation}
The gauge fields and gauge parameters are normalized as
\begin{align}
& 
\int_{\cal C} d\Phi_0^{\rm M},
\,
 \int_{\cal S} dA_1^{\rm M},
\,
 \int_{\cal V} dB_2^{\rm M},
\,
 \int_{\Omega} dC_3^{\rm M} \in 2\pi \bb{Z},
\label{210515.0248}
\\
& 
\int_{\cal C} d\Lambda_0^{\rm M},
 \int_{\cal S} d\Lambda_1^{\rm M},
 \int_{\cal V} d\Lambda_2^{\rm M}
 \in 2\pi \bb{Z},
\end{align}
respectively.
Under the normalization, one can show that the action 
$S_{\rm bg,M} [X_5]$ mod $2\pi$ does not depend on a choice of the
auxiliary space $X_5$.

The simultaneous gauging of the electric and magnetic symmetries can be
done by the coupling of the electric background gauge fields to $S_{\rm
bg,M}$.
The total background gauged action is 
\begin{equation}
\begin{split}
    S_{\rm EM bg} [X_5]
&
= 
\int_{X_5} \Big( \fr{k}{2\pi}
(dc - D^K_4) 
\wed  (d\phi -A^m_1) 
+ \fr{q}{2\pi} (db - C^Q_3) \wed (da - B^p_2)
\\
&
\quad
\hph{\int_{X_5} \Big( }
+
\fr{N}{8\pi^2} (d\phi - A^m_1) \wed (da - B^p_2) \wed (da -B^p_2)\Big) 
\\
&
\quad
+
\fr{1}{2\pi}\int_{X_5}
 \Big((dc - D_4^K) \wed d\Phi_0^{\rm M}
-
 (db - C_3^Q) \wed d A_1^{\rm M} 
\\
&
\quad
\hph{
+
\fr{1}{2\pi}\int_{X_5}
 \Big(}
+
 (da - B_2^p) \wed dB_2^{\rm M} 
-
(d\phi - A_1^m)  \wed dC_3^{\rm M}\Big)
. 
\end{split}
\label{210515.0056}
\end{equation} 
Since $dB_2^{\rm M}$ and 
$qC_3^Q= dC_2^q +\fr{N}{2\pi } A_1^m \wed B_2^p $ 
are coupled to the same current $\frac{1}{2\pi}(da-B_2^p)$, we can
absorb the $dB_2^{\rm M}$ by shifting $C_2^q\to C_2^q-B_2^{\rm M}$ while
preserving flux quantization conditions in \ers{210515.0211} and
\eqref{210515.0248}.
Similarly, we can also absorb $dC_3^{\rm M}$ by the shift 
$D_3^k \to D_3^k-C_3^{\rm M}$.
Therefore, the background gauging of magnetic $U(1)$ $1$- and $2$-form
symmetries are redundant if we treat $C_3^Q$, $C_2^q$, $D_4^K$, and
$D_3^k$ as independent gauge fields.
On the other hand, $d\Phi_0^{\rm M}$ and $dA_1^{\rm M}$ are coupled to
currents different from those of $A_1^m$ and $B_2^p$, so that
$d\Phi_0^{\rm M}$ and $dA_1^{\rm M}$ cannot be absorbed by the shift of
gauge fields.
The resultant gauged action is 
\begin{equation}
\begin{split}
    S_{\rm TAE, EM bg} [X_5]
&
= 
\int_{X_5} \Big( \fr{k}{2\pi}
(dc - D^K_4) 
\wed  (d\phi -A^m_1) 
+ \fr{q}{2\pi} (db - C^Q_3) \wed (da - B^p_2)
\\
&
\quad
\hph{\int_{X_5} \Big( }
+
\fr{N}{8\pi^2} (d\phi - A^m_1) \wed (da - B^p_2) \wed (da -B^p_2)\Big) 
\\
&
\quad
+
\fr{1}{2\pi}\int_{X_5}
 \Big((dc - D_4^K) \wed d\Phi_0^{\rm M}
-
 (db - C_3^Q) \wed d A_1^{\rm M} 
 \Bigr). 
\end{split}
\end{equation}

\subsection{'t Hooft anomalies}
We have obtained the gauged action consistent with the gauge invariance
of the dynamical fields.
Meanwhile, we have the ambiguity due to only the background gauge fields,
\begin{equation}
\begin{split}
&
\int_{Z_5} \Big( \fr{k}{2\pi}
\(dc - D^K_4\) 
\wed  (d\phi -A^m_1) 
+ \fr{q}{2\pi} (db - C^Q_3) \wed (da - B^p_2)
\\
&
\quad
\hph{\int_{X_5} \Big( }
+
\fr{N}{8\pi^2} (d\phi - A^m_1) \wed (da - B^p_2) \wed (da -B^p_2)\Big) 
\\
&
+\fr{1}{2\pi}\int_{Z_5}
 \Big((dc - D_4^k) \wed d\Phi_0^{\rm M}
-
 (db - C_3^q) \wed d A_1^{\rm M} 
\\
&
\hph{
+
\fr{1}{2\pi}\int_{X_5}
 \Big(}
+
 (da - B_2^p) \wed dB_2^{\rm M} 
-
(d\phi - A_1^m)  \wed dC_3^{\rm M}\Big)
\\
&
= 
\int_{Z_5} \Big( \fr{k}{2\pi}
D^k_4
\wed A^m_1 
+ \fr{q}{2\pi} C^q_3 \wed  B^p_2
+ \fr{N}{4\pi^2} A_1^m \wed B_2^p \wed B^p_2
\Big)
\\
&
\quad
+
\fr{1}{2\pi}\int_{Z_5}
 \Big(
- D_4^K \wed d\Phi_0^{\rm M}
+ 
 C_3^Q \wed d A_1^{\rm M}
\Big).
\end{split}
\label{210515.0026}
\end{equation} 
This is an 't Hooft anomaly, which is an obstruction to 
gauging global symmetries dynamically.
In our case, the term with fractional number
$\fr{k}{2\pi} \int_{Z_5} D^k_4 \wed A^m_1 \in \fr{2\pi}{m} \bb{Z}$ 
implies that 
we cannot gauge the pair of the 
0- and 3-form symmetries.
Similarly, the presence of 
$\fr{q}{2\pi} \int_{Z_5} C^q_3 \wed  B^p_2 \in \fr{2\pi}{p}\bb{Z}$
prevents us from gauging the pair of 1- and 2-form symmetries.

Furthermore, the gauging of the pair of 0- and 1-form symmetries 
is forbidden in the presence of 
$\fr{N}{4\pi^2} \int_{Z_5} A_1^m \wed B_2^p \wed B^p_2 \in \fr{4 \pi N}{mp^2}\bb{Z} $. 
This type of anomaly is called the 2-group anomaly~\cite{Benini:2018reh}.
We also have 't Hooft anomalies due to the simultaneous gauging of the electric and magnetic symmetries.
Both of the two terms represent the 
mixed 't Hooft anomalies which forbid the dynamical gauging of 
the electric and magnetic symmetries associated with 
the equations of motion and Bianchi identities for the 
dynamical fields.
In the presence of the 't Hooft anomalies,
a symmetry preserving gapped vacuum is forbidden.
This is consistent with the fact that 
the axion has $m$ of degenerated vacua
connected by domain walls $U_0$, and the photon is in a 
topologically ordered phase as we discuss in section \ref{BGphys}.

The 't Hooft anomalies can also be seen in a viewpoint of 
a 5-dimensional theory as follows.
We consider the following topological action,
\begin{equation}
\begin{split}
S_{\rm 5D} [Z_5]
&
= 
\int_{Z_5} \Big( \fr{k}{2\pi}
D^K_4
\wed A^m_1 
+ \fr{q}{2\pi} C^Q_3 \wed  B^p_2
- \fr{N}{8\pi^2} A_1^m \wed B_2^p \wed B^p_2
\Big)
\\
&
\quad
+
\fr{1}{2\pi}\int_{Z_5}
 \Big(
- D_4^K \wed d\Phi_0^{\rm M}
+ 
 C_3^Q \wed d A_1^{\rm M}
\Big),
\end{split}
\label{210703.1828}
\end{equation} 
which is gauge invariant if $Z_5$ does not have boundaries.
If $Z_5$ has boundaries, the gauge invariance is violated
at the boundaries.
The violation of the gauge invariance matches the 't Hooft anomalies 
in the topological axion electrodynamics.
This means that the 't Hooft anomalies in the topological 
axion electrodynamics can be canceled by $S_{\rm 5D}$
via the anomaly inflow mechanism~\cite{Callan:1984sa}.

\subsection{Modified gauge fields as higher-group gauge fields}
\label{sec:higher group structure}
We here discuss the underlying mathematical structure  
for the modifications of the background gauge fields.
Following Ref.~\cite{Cordova:2018cvg},
 we refer to a set of $0$-,..., $(n-1)$-form 
symmetry groups, $\{G_0,\cdots, G_{n-1}\}$, with nontrivial correlations as an $n$-group. 
In the context of physics, 
it will be clearer to express the $n$-group structure
in terms of gauge theory:
we refer to a set of $1$-,..., $n$-form gauge fields 
with mixed gauge transformation laws 
as an $n$-group gauge theory.
Therefore, the $n$-group can be characterized as
a set of groups 
whose gauge theory organizes an $n$-group gauge theory.

In our case, 
we can argue that 
the higher-form symmetries of the topological axion electrodynamics 
organize a 4-group, 
since we have the 1-,..., 4-form background gauge fields for 
the global symmetries, which have mixed gauge transformation laws.
Furthermore, we can specify a detailed mathematical structure of the 4-group by the field strengths as follows.

The key equations
\eqref{eq: modified gauge fields1},
and \eqref{eq: modified gauge fields2} 
in addition to $dB_1^p=pB_2^p$
can be 
expressed as%
\footnote{The conditions in \er{210824.1622} 
are called vanishing fake curvature conditions~\cite{Baez:2004in,Baez:2005qu,Baez:2010ya}.}
\begin{align}
    dB^p_1 =\partial_1 B^p_2,\quad
    dC^q_2 + A_1^m\triangleright  C_2^q=\partial_2 C_3^Q,\quad
    dD^k_3 + \{B^p_2,B^p_2\}=\partial_3 D_4^K,
    \label{210824.1622}
\end{align}
where we have defined
\begin{align}
&\partial_1 B_2^p = p B_2^p,\quad
\partial_2 C_3^Q = q C_3^Q,\quad
\partial_3 D_4^K = k D_4^K, \\
&A_1^m\triangleright C_2^q = \frac{N}{2\pi}A_1^m\wedge B_2^p, \quad
\{B_2^p,B_2^p\} = \frac{N}{4\pi}B_2^p\wedge B_2^p.
\end{align}
From these data, we find that our 4-group can be classified into a
so-called semi-strict 4-group or 3-crossed module denoted as
$(G_3\overset{\partial_3}{\to} G_2\overset{\partial_2}{\to} G_1\overset{\partial_1}{\to} G_0
, \triangleright, \{-,-\})$ 
in the mathematical literature~\cite{Arvasi:2009}.
We explain the ingredients of the semi-strict 4-group
as follows:%
\footnote{We may include $G_{-1} = U(1)$
whose gauge field is $A_0^m$.
In this case, the boundary map $\der_0 : G_0 \to G_{-1}$ 
is given by 
$\der_0 (e^{2\pi i n_0 /m}, e^{i\theta_0})  = e^{m \cdot 2\pi i n_0/ m} =1$.}
\begin{enumerate}
    \item 
$G_n$ are groups, which are  
$G_0=\mathbb{Z}_m\times U(1)$, 
$G_1=\mathbb{Z}_p\times U(1)$,
$G_2=\mathbb{Z}_Q\times U(1)$, 
and $G_3=\mathbb{Z}_K$. 
The corresponding gauge fields are 
$(A_1^m, B_1^p)$, $(B_2^p,C_2^q)$, $(C_3^Q,D_3^k)$, and $D_4^K$, 
with mixed gauge transformation laws
given by \ers{210515.0144}, \eqref{210515.0145}, \eqref{210515.0146},
\eqref{210515.0147},
\eqref{210703.2232}, and \eqref{210703.2233}.

\item 
Boundary maps $\partial_n: G_n\to G_{n-1}$ are maps from
 electric symmetries to magnetic symmetries, which satisfy
$\partial_{n-1}\circ \partial_n (g_n) =1 \in G_{n-2}$
for $g_n \in G_n$.
Concretely, for 
$(e^{2\pi in_2/Q},e^{i\theta_2})\in\mathbb{Z}_Q\times U(1)=G_2$, 
and $e^{2\pi in_3/K}\in\mathbb{Z}_K=G_3$, the maps are
\begin{align}
    \partial_2(e^{2\pi in_2/Q},e^{i\theta_2})
 &
=  (1,e^{2\pi in_2 q/Q}) \in \mathbb{Z}_p \times U(1) = G_1,
\\
    \partial_3 e^{2\pi in_3/K}
&
=  (1,e^{2\pi in_3k/K})
    \in \mathbb{Z}_Q \times U(1) = G_2.
\end{align}
In contrast, $\partial_1$ has no nontrivial structure 
because it sends the element of $\mathbb{Z}_p$ to the identity element: 
$\mathbb{Z}_p \times U(1) \ni (e^{2\pi in_1/p}, e^{i\theta_1})\mapsto 
(1,e^{2\pi in_1})=(1,1)\in \bb{Z}_m \times U(1)$.
Kernels of $\partial_n$ represent the groups of 
higher-form global symmetries.
In particular, the kernels of $\partial_2$ and $\der_3$, 
$ \mathrm{ker} \, \partial_2 = \mathbb{Z}_q \times U(1) \subset G_2$ 
and 
${\rm ker}\, \der_3 = \bb{Z}_k \subset G_3$,
are the groups of 2- and 3-form symmetries,
respectively.

\item 
There is a group action of $G_0$ on $G_n$ denoted by $\triangleright$.
In our model, the only action on $G_1$ is nontrivial.
In concrete, for $(e^{2\pi in_0 /m},e^{i\theta_0})\in \mathbb{Z}_m\times U(1)=G_0$, and $(e^{2\pi i n_1/p},e^{i\theta_1})\in \mathbb{Z}_p \times U(1) =G_1$,
the action of $G_0$ is 
\begin{equation}
(e^{2\pi in_0 /m},e^{i\theta_0})
\triangleright
(e^{2\pi i n_1/p},e^{i\theta_1}) 
= 
(e^{2\pi i n_1/p},e^{2\pi i n_0n_1N/(mp)}e^{i\theta_1}).
\end{equation}

\item 
A map $\{-,-\}: G_1\times G_1 \to G_2$ is called 
the Peiffer lifting~\cite{CONDUCHE1984155}. 
For elements of $G_1$, 
$(e^{2\pi i n_1/p},e^{i\theta_1}),(e^{2\pi i n'_1/p},e^{i\theta'_1})\in \mathbb{Z}_p \times U(1)=G_1$,
the Peiffer lifting reads
\begin{equation}
\{(e^{2\pi in_1/p},e^{i\theta_1}),(e^{2\pi in'_1/p},e^{i\theta'_1})\}
=
(1,e^{2\pi i n_1 n'_1 N/(2p^2)})
\in \mathbb{Z}_Q \times U(1)=G_2.
\end{equation}
\item 
Operators $\{-,-\}$, $\triangleright$, and $\partial_n$ satisfy
several consistency conditions (axioms).
For example, the group action $\triangleright$ is consistent with the
Peiffer lifting:
\begin{equation}
    g\triangleright\{h_1,h_2\}= \{g\triangleright h_1,g\triangleright h_2\},
\end{equation}
where $g\in G_0$, and $h_1,h_2\in G_1$.
The consistency conditions mean that the symmetry generators do not depend on the order of deformation.
\end{enumerate}

Note that we can check that our 4-group satisfies 
the axioms given in Ref.~\cite{Arvasi:2009} by using the above definitions.
This class of the higher-groups 
has been found in the context of the quantum 
chromodynamics~\cite{Tanizaki:2019rbk},
where a field strength of a 3-form gauge field is modified
by a wedge product of a flat 2-form gauge field
and a 4-form gauge field
similar to \er{eq: modified gauge fields2}.%
\footnote{Note that the structure 
of the 4-group in Ref.~\cite{Tanizaki:2019rbk} can be identified 
as a semi-strict 4-group 
$(G_3 \overset{\der_3}{\to} G_2 \to G_1 \to G_0, \{-,-\}) $,
where $G_3 = \bb{Z}_{Np}$, $G_2 = U(1)$, $G_1 = \bb{Z}_N$, $G_0 = 1$,
$\der_3 e^{2\pi i n_3/ (Np)} = e^{2\pi i n_3/N}$,
and  
$\{e^{2\pi i n_1/N}, e^{2\pi i n_1'/N}\} = e^{2\pi i n_1n_1'/2N} \in G_2$.
This group structure can be derived by 
the modified field strength of a 3-form gauge field,
$dD^p_3 + \fr{N}{4\pi} B^N_2 \wed B^N_2 = pD^p_4$
in our notation. 
}
Meanwhile, the 4-group structure discussed in our paper may be the first example where all of the 0-,..., 3-form 
symmetry groups are nontrivially correlated.

We comment on a physical interpretation of semi-strict higher-groups. 
In the language of gauge theories for the semi-strict higher-groups,
the field strengths 
of higher-form gauge fields are modified by 
quadratic forms of 
lower-form gauge fields~\cite{Baez:2002jn, Baez:2004in, Baez:2005qu, Martins:2009evc, Wang:2013dwa, Saemann:2013pca}.
Physically, there are boundaries of symmetry generators 
on intersections of two symmetry generators,
since the field strengths of the background gauge fields specify
configurations of boundaries of the symmetry generators.
These modifications can be understood as natural extensions of 
non-Abelian gauge theories of ordinary non-Abelian groups,
where the field strengths should have the quadratic terms of gauge fields
if the structure constants are non-zero.

More generally, higher-groups can be weak: 
field strengths of higher-form gauge fields 
are modified by field strengths or cubic (or higher) forms
of lower-form gauge fields~\cite{Cordova:2018cvg}.
In particular, structures of weak 2-groups
have been investigated in detail, where the modifications 
of field strengths of 2-form gauge fields are given by 
Chern-Simons forms or Postnikov classes~\cite{Baez:2003fs, Kapustin:2013uxa, Sharpe:2015mja, Benini:2018reh}.
The modifications can be understood as generalizations of 
2-form gauge fields
in the heterotic string theories, 
whose field strengths are modified by the Chern-Simons terms 
of the Yang-Mills and local Lorentz gauge fields
via the Green-Schwarz mechanism~\cite{Green:1984sg}.

\section{\label{BGphys}Physical effects in topological axion electrodynamics}
In this section, we discuss some physical effects in the topological
axion electrodynamics by using both the background gauged actions and
correlation functions of symmetry generators.

\subsection{Topological order in bulk}

Here, we argue that the topological axion electrodynamics 
in the bulk exhibits an Abelian type of  topological order for $p = \gcd(q,N) \neq 1$.
In particular, we show that the fractional statistics 
is given by $\bb{Z}_p$,
which is in contrast to the fractional phase in the ordinary Abelian Higgs model whose fractional statistics is given by the charge of the Higgs field, $\bb{Z}_q$.

In terms of the background gauging, the existence of the topological order 
can be directly seen by the topological term 
$\int_{Z_5} \fr{q}{2\pi } B_2^p \wed C_3^q \in \fr{2\pi}{p} \bb{Z}$ 
in \er{210515.0026},
which expresses 
the mixed 't Hooft anomaly between 
1- and 2-form symmetries.
Since the 't Hooft anomaly is $\bb{Z}_p$-valued, 
we conclude that the ground state has  
degeneracy 
classified by the configurations of $B_2^p$ and $C_3^q$
as well as 
topology of a 
spatial manifold.
For example, if the spatial manifold
is $S^2 \times S^1$, we have $p$-fold 
degeneracy~(see, e.g., Ref.~\cite{Hidaka:2019jtv} in detail). 

The discussion based on the 't Hooft anomaly is direct and straightforward, but it may not be physically intuitive.
In the following, we explain the topological order in terms of symmetry generators, which will be more intuitive than the above argument.

\subsubsection{Non-local order parameters and fractional linking statistics}

In order to find the topological order in $(3+1)$ dimensions, 
we should find non-local order parameters,
which are topological and have fractional linking statistics.
We will call them topological order parameters.
Since symmetry generators are non-local and 
topological, they are candidates for 
the order parameters.
In the following, we show
that the symmetry generators $U_1$ and $U_2$ can be regarded as 
topological order parameters.

The topological order can be characterized by the following 
correlation function which can be evaluated by the same procedure
summarized in Appendix~\ref{corr},
\begin{equation}
\begin{split}
 \vevs{U_1 (e^{2\pi i n_1/p}, {\cal S}) U_2 (e^{2\pi i n_2/q}, {\cal C})}
&=
e^{- 2\pi i \fr{n_1 n_2 }{p} \link ({\cal S,C})} 
\vevs{ U_2 (e^{2\pi i n_2/q}, {\cal C})}
\\
&=
e^{- 2\pi i \fr{n_1 n_2 }{q} \cdot \fr{q}{p} \link ({\cal S,C})} 
\vevs{ U_1 (e^{2\pi i n_1/p}, {\cal S})}
\\
&=
e^{- 2\pi i \fr{n_1 n_2 }{p} \link ({\cal S,C})} .
\end{split}
\label{210516.1648}
\end{equation}

We explain the physical meanings of \er{210516.1648}.
The right-hand side of the first line 
shows that $U_2$ is charged under the action of $U_1$
with the charge $-n_2$.
The second line implies that $U_1$ is also charged 
under $U_2$ with the charge $-n_1 q / p$.
Note that $U_1$ belongs to the representation of $\bb{Z}_q$
parameterized by $\bb{Z}_p = \bb{Z}_{\gcd (N,q)}$
while the symmetry generator $U_2$ is parameterized by 
the group $\bb{Z}_q$.
The third line means that the symmetry generators 
$U_1$ and $U_2$ has a fractional linking phase.
It is the AB effect with a fractional phase:
an electrically charged test particle
receives a fractional phase when it encircles 
a string-like quantized magnetic field.

Since the topological order parameters develop non-zero VEVs
$\vevs{ U_2 (e^{2\pi i n_2/q}, {\cal C})} =
\vevs{ U_1 (e^{2\pi i n_1/p}, {\cal S})} = 1$ 
and they have fractional linking phases, the topological axion electrodynamics
 is topologically ordered. 
The symmetry generators consist of groups so that there is no nontrivial fusion rule, which implies this is an Abelian type of topological order.
This topologically ordered phase can be understood as 
a symmetry broken phase of both of the $\bb{Z}_p$ 1-form and $\bb{Z}_q$ 
2-form symmetries, since the charged objects develop non-zero VEVs.
Furthermore, the symmetry breaking pattern can be classified 
as the type-B spontaneous symmetry breaking,
since the charged objects are symmetry generators~\cite{Nielsen:1975hm,Nambu:2004yia,Watanabe:2011ec,Watanabe:2012hr,Hidaka:2012ym}.

\subsubsection{Comparison to topological order in Abelian Higgs model}
Here, we discuss the difference of the topologically ordered phases 
between the topological axion electrodynamics and Abelian Higgs models.
The Abelian Higgs model 
with a charge $q$ Higgs field
can be topologically ordered in the low-energy limit~\cite{Hansson:2004wca}.
On the one hand, 
the fractional linking phase
is determined by the charge of the Higgs field
as $\bb{Z}_q$ for the Abelian Higgs model.
On the other hand, the linking phase is deformed by 
the axion-photon coupling as $\bb{Z}_p = \bb{Z}_{\gcd(N,q)}$
for the topological axion electrodynamics.
Therefore the global 1-form symmetries are different between the topological axion electrodynamics and Abelian Higgs models.
Physically, the axion and Higgs fields screen $N$ and $q$ of quantized magnetic fields, respectively.

\subsection{Topological order on axionic domain wall}
Next, we consider the topological order 
on the axionic domain wall in the viewpoint 
of the background gauge field.
The nontrivial ordered phase corresponds to 
the 't Hooft anomaly 
$\fr{N}{8\pi^2} \int_{Z_5} A_1^m \wed B_2^p \wed B_2^p$ in \er{210703.1828}.
The topological term in five dimensions means 
that the 1-form symmetry generators have nontrivial linking phase
($\sim B_2^p \wed B_2^p$) on a worldvolume 
of the axionic domain wall represented by 
$A_1^m$.
The 't Hooft anomaly implies that  
the ground state in the existence of the domain wall is not uniquely gapped.
By the fractional phases of the flat gauge fields,
the ground state exhibits the topological order characterized by $\bb{Z}_{P}$ group,
where 
$P\coloneqq mp^2/\gcd (N, mp^2)$ is the nontrivial 
denominator of $N/ (mp^2)$.

\subsubsection{Intersection of 0- and 1-form symmetry generators}
\label{sec:Intersection of 0- and 1-form symmetry generators}
In the following, we give a detailed review 
on the intersection of symmetry generators to
discuss the topological order on the axionic domain wall~\cite{Hidaka:2021mml}.
In order to show the topological order, 
we need to intersect the symmetry generators 
$U_0$ and $U_1$.
As we will see below, we should carefully treat 
the intersection of the symmetry generators.

First, we naively consider a correlation function of 0- and 1-form symmetry generators with intersections,
$
\vevs{
U_0 (e^{2\pi i n_0/m}, {\cal V}) 
U_1 (e^{2\pi i n_1/p}, {\cal S}_1)
}$
where ${\cal V}$ and ${\cal S}_1$ 
are 3- and 2-dimensional closed subspace 
without self-intersections.
We assume that 
${\cal V} \cap {\cal S}_1$ is a closed 1-dimensional subspace.
We can evaluate a correlation function  
absorbing $U_1$ and $U_0$ to the action 
by the redefinition of $a$ and $\phi$ as
\begin{equation}
    \vevs{
U_0 (e^{2\pi i n_0/m}, {\cal V}) 
U_1 (e^{2\pi i n_1/p}, {\cal S}_1)
}
= 
\vevs{
e^{-2\pi i \fr{N n_0 n_1}{mp} 
\int_{\Omega_{\cal V}} \frac{da}{2\pi} \wed \delta_2 ({\cal S}_1)
}
}.
\end{equation}
However, the object on the right-hand side may violate the large gauge 
invariance of the photon, if 
the coefficient $Nn_0 n_1/ mp$ is fractional.
The violation of the large gauge invariance 
can be shown by the ambiguity of the 
choice of $\Omega_{\cal V}$.
In the above correlation function, we can choose another 
4-dimensional subspace $\Omega'_{\cal V}$ whose boundary is ${\cal V}$.
Since the left-hand side of the correlation function does not depend on the choice, the right-hand side should also be independent of the choice. 
However, when we replace the 4-dimensional manifold, 
we have an additional phase
$ e^{-2\pi i \fr{N n_0 n_1}{mp} 
\int_{\Omega} \frac{da}{2\pi} \wed \delta_2 ({\cal S}_1)}
$,
\begin{equation}
 e^{-2\pi i \fr{N n_0 n_1}{mp} 
\int_{\Omega_{\cal V}} \frac{da}{2\pi} \wed \delta_2 ({\cal S}_1)}
 = 
 e^{-2\pi i \fr{N n_0 n_1}{mp} 
\int_{\Omega} \frac{da}{2\pi} \wed \delta_2 ({\cal S}_1)}
 e^{-2\pi i \fr{N n_0 n_1}{mp} 
\int_{\Omega_{\cal V}'} \frac{da}{2\pi} \wed \delta_2 ({\cal S}_1)},
\end{equation}
and the phase can be nontrivial if we include an 't Hooft line 
in the correlation function.
Therefore, we carefully treat intersections of symmetry generators with respect to the large gauge invariance.

\begin{figure}[t]
\begin{center}
\ig[height=7em]{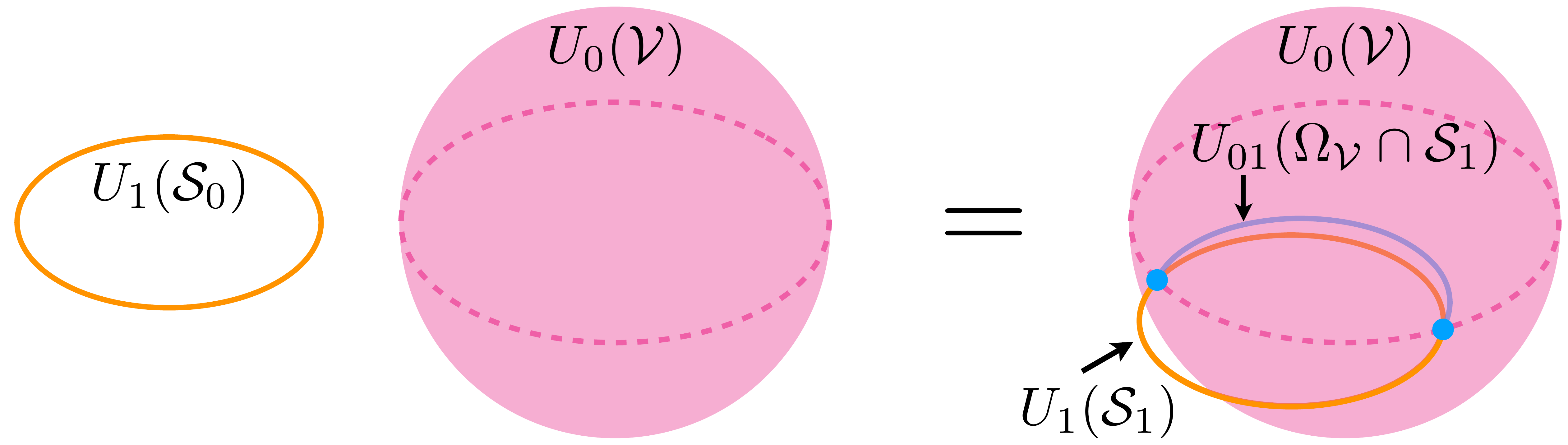}
\end{center}
\caption{\label{U01}
The intersection of 
0- and 1-form symmetry generators.
This figure shows a time slice of the symmetry generators.
The 0- and 1-form symmetry generators are introduced on 3- and 2-dimensional subspaces, which are temporally and spatially extended.
The pink sphere and orange line correspond
to the 0- and 1-form symmetry generators on 
the time slice, respectively.
The blue line in the right panel is a time slice of an induced static surface $U_{01}$.
The blue dots denote a time slice of a temporally extended loop, which is the boundary of $\Omega_{{\cal V}}\cap {\cal S}_1$ on the time slice,
which corresponds to induced anyons on the domain wall.
We have abbreviated the parameters of the symmetry generators for simplicity.
}
\end{figure}

In order to discuss the intersection carefully,
we take two symmetry generators 
$U_0 (e^{2\pi i n_0/m}, {\cal V}) $ 
and 
$U_1 (e^{2\pi i n_1/p}, {\cal S}_0)$
where 
we assume that two symmetry generators 
are not intersected with each other,
${\cal V} \cap {\cal S}_0 = \emptyset$
(see Fig.~\ref{U01}).
We also assume that ${\cal S}_0$ does not 
have any self-intersections.
Since each of the symmetry generators is 
contractible, the correlation function given by the two symmetry generators becomes trivial:
\begin{equation}
\vevs{
U_0 (e^{2\pi i n_0/m}, {\cal V}) 
U_1 (e^{2\pi i n_1/p}, {\cal S}_0)
}=1.
\end{equation}
We now intersect them by deforming the worldsheet 
${\cal S}_0 $ to ${\cal S}_1$
with the condition 
${\cal S}_0 \cap {\cal S}_1 = \emptyset$.
Here, we assume that the intersection
${\cal V} \cap {\cal S}_1$ 
is a 1-dimensional closed subspace.
The deformation can be done by 
interpolating them with 
a 3-dimensional subspace ${\cal V}_{01}$
satisfying 
$\der {\cal V}_{01} =  {\cal S}_0 \cup \b{\cal S}_1$. 
We also assume that ${\cal V}_{01}$ does not intersect with 
any singularity such as an 't Hooft line.
Under the deformation, 
we can rewrite the 1-form symmetry generator 
as $U_1 (e^{2\pi i n_1/p}, {\cal S}_0)
= 
U_1 (e^{2\pi i n_1/p},  \der {\cal V}_{01} )
U_1 (e^{2\pi i n_1/p}, {\cal S}_1).
$
The correlation function can be rewritten as \begin{equation}
\begin{split}
& \vevs{U_0 (e^{2\pi i n_0/m}, {\cal V}_0) 
U_1 (e^{2\pi i n_1/p}, {\cal S}_0)}
\\
&
\quad= 
 \vevs{U_0 (e^{2\pi i n_0/m}, {\cal V}) 
U_1 (e^{2\pi i n_1/p},  \der {\cal V}_{01} )
U_1 (e^{2\pi i n_1/p}, {\cal S}_1)}.
\end{split}
\end{equation}
The symmetry generator $U_1 (e^{2\pi i n_1/p},  \der {\cal V}_{01} )$
 can be absorbed to the action by 
the redefinition $a + \fr{2\pi n_1}{p} \delta_1 ({\cal V}_{01}) \to a$
as
\begin{equation}
\begin{split}
&\vevs{U_0 (e^{2\pi i n_0/m}, {\cal V}) 
U_1 (e^{2\pi i n_1/p}, {\cal S}_0)}
\\
&
\quad= 
 \vevs{
e^{
2\pi i\fr{N}{mp}n_0 n_1 
\int_{\Omega_{\cal V}} 
\frac{d a}{2\pi}
\wed 
\delta_2 ({\cal S}_1)
}
U_0 (e^{2\pi i n_0 / m}, {\cal V}) 
U_1 (e^{2\pi i n_1 / p}, {\cal S}_1)
},
\end{split}
\label{210708.2321}
\end{equation}
where $\Omega_{\cal V}$ is a 4-dimensional subspace 
whose boundary is ${\cal V}$,
and we have used 
$d\delta_1 ({\cal V}_{01})
 = - \delta_2({\cal S}_0) + \delta_2 ({\cal S}_1)$.
Since 
$\Omega_{\cal V} \cap {\cal S}_1$
is a 2-dimensional subspace whose boundary is
${\cal V} \cap {\cal S}_1$, 
we have a 2-dimensional object with the boundary, 
\begin{equation}
U_{01} (e^{2 \pi i \fr{N}{mp}n_0n_1}, 
\Omega_{\cal V} \cap {\cal S}_1)
=
e^{2\pi i\fr{N}{mp}n_0n_1  \int_{\Omega_{\cal V}} \frac{da}{2\pi} \wed 
\delta_2 ({\cal S}_1)}.
\end{equation}
Therefore, we should have an additional object on a 2-dimensional 
subspace if we try to intersect them.
Since the correlation function in \er{210708.2321}
is trivial, there should be an electrically charged object with a fractional charge 
$Nn_0n_1/ (mp)$.

Physically, the fractional charge on the intersection means the Sikivie effect
and anomalous Hall effect.
If we take ${\cal S}_1$ as 
a spatially and temporally extended object, 
the symmetry generator represents 
a worldsheet of a quantized magnetic flux.
The Sikivie effect implies that 
there is an induced electric charge 
on the intersection of the axionic domain wall
and the magnetic flux~\cite{Sikivie:1984yz}.
If we instead take ${\cal S}_1$ as 
an instantaneous surface, the 1-form symmetry generator can be understood as an 
external electric field.
The anomalous Hall effect implies that 
there is an induced electric current 
on the axionic domain wall~\cite{Sikivie:1984yz, Wilczek:1987mv, Essin:2008rq, Qi:2008ew}.
Since the axionic domain wall can be 
understood as a fractional quantum Hall system
because of the Chern-Simons term in $U_0$,
the induced electric charge or current 
can be identified as an anyon.

The necessity of the additional object $U_{01}$ can be naturally understood as a natural consequence of the deformation of 
the 3-form gauge field $C_3^Q$ in \er{210709.1701}.
Since $A_1^m \wed B_2^p$ implies the intersection of the 0- and 1-form symmetry generators, 
the modification in \er{210709.1701} means that there should be a 2-form symmetry generator
on the intersection.
Since \er{210708.2321} is trivial, 
we have an object canceling $U_{01}$,
which can be identified as 
a 2-form symmetry 
generator by the Stokes theorem.

\subsubsection{Intersection of two 1-form symmetry generators}

In order to discuss the topological order on the domain walls,
we need to consider a link of anyons.
This configuration can be constructed by 
using 1-form symmetry generators,
which are intersected with each other in the bulk.
As in the above discussion, 
we take two 1-form symmetry generators 
$U_1 (e^{2\pi i n_1/p}, {\cal S})$
and 
$U_1 (e^{2\pi i n'_1/p}, {\cal S}'_0)$,
which are not intersected,
${\cal S} \cap {\cal S}_0' =\emptyset$
and do not have any self-intersection
(see Fig.~\ref{U11}).
{\small
\begin{figure}[t]
\begin{center}
\ig[height=7em]{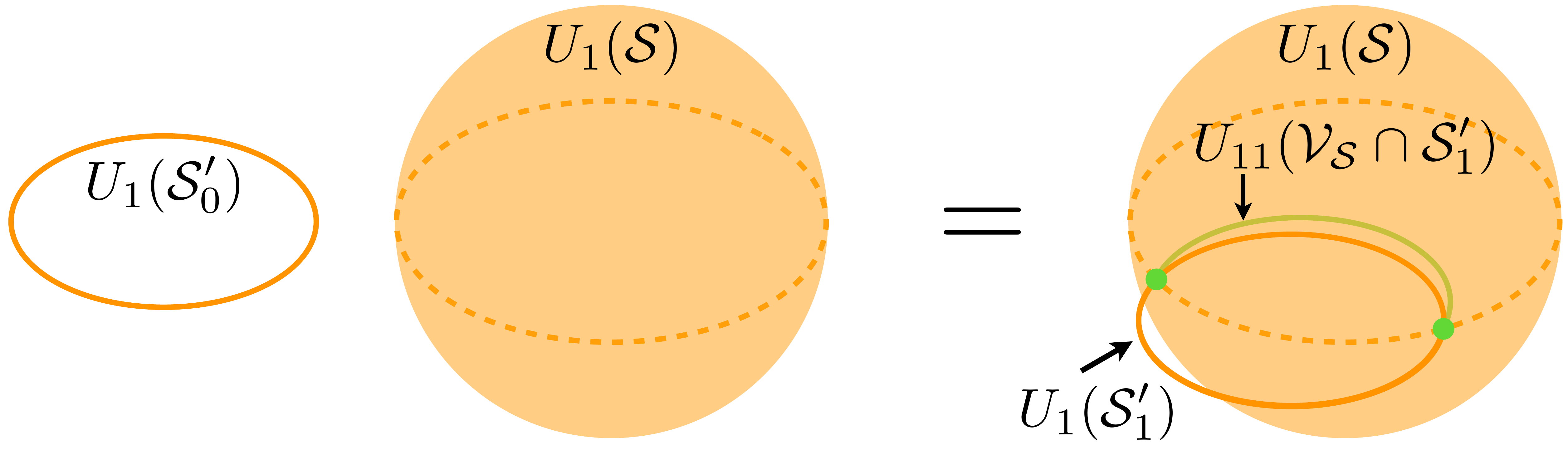}
\end{center}
\caption{\label{U11}
The intersection of two 1-form symmetry generators at a time slice.
The 1-form symmetry generators are
introduced 
on 2-dimensional subspaces. 
One of them 
is given as an instantaneous 
sphere ${\cal S}$, which is 
described by 
an orange sphere in the figure.
The other is given as 
a temporally and spatially 
extended 2-dimensional closed subspace ${\cal S}_{0,1}'$, 
which can be seen as 
a circle on the time slice 
as shown in an orange circle
in the figure.
The green line in the right panel is a time slice of an induced surface $U_{11}$, which is extended to spatial and temporal directions.
The green dots on the boundaries 
of the green line are
instantaneous objects representing 
an induced 3-form symmetry generator.
The dots can be physically interpreted as induced axions.
We have again abbreviated the parameters of the symmetry generators.
}
\end{figure}
}
The correlation function of two symmetry 
generators is trivial,
since both of them can be continuously contracted,
\begin{equation}
\begin{split}
 \vevs{
U_1 (e^{2\pi i n_1/p}, {\cal S})
U_1 (e^{2\pi i n'_1/p}, {\cal S}'_0)
}=1.
\end{split}
\end{equation}
Now, we deform ${\cal S}'_0$ to ${\cal S}'_1$ 
that is intersected with ${\cal S}$. 
We can deform it by interpolating with 
a 3-dimensional subspace 
${\cal V}'_{01}$
satisfying 
 $\der {\cal V}'_{01} = 
 {\cal S}'_0\cup \b{\cal S}'_1$.
By using 
$U_1 (e^{2\pi i n'_1/p}, {\cal S}'_0)
 = U_1 (e^{2\pi i n'_1/p}, \der {\cal V}'_{01})
U_1 (e^{2\pi i n'_1/p}, {\cal S}'_1)$ 
and by absorbing 
$U_1 (e^{2\pi i n'_1/p}, \der {\cal V}'_{01})$
into the action,
we obtain
\begin{equation}
\begin{split}
&\vevs{
U_1 (e^{2\pi i n_1/p}, {\cal S})
U_1 (e^{2\pi i n'_1/p}, {\cal S}'_0)}
\\
&\quad =  
\vevs{
e^{2\pi i \fr{N}{p^2} n_1n_1' 
\int_{{\cal V_S}}
 \frac{d\phi}{2\pi} \wed \delta_2 ({\cal S}'_1)}
U_1 (e^{2\pi i n_1/p}, {\cal S})
U_1 (e^{2\pi i n'_1/p}, {\cal S}'_1)
}
.
\end{split}
\end{equation}
Here, ${\cal V_S}$ is a 3-dimensional subspace 
whose boundary is ${\cal S}$, 
and we have used 
${\cal V_S} \cap \der {\cal V}_{01}' = 
{\cal V_S} \cap \b{\cal S}_1'$.
We thus obtain an object,
\begin{equation}
U_{11} (e^{2 \pi i \fr{N}{p^2} n_1n_1' }, 
{\cal V_S}\cap {\cal S}_1')
\coloneqq
e^{2\pi i \fr{N}{p^2} n_1n_1' 
\int_{\cal V_S}
 \frac{d\phi}{2\pi} \wed \delta_2 ({\cal S}_1')}
\end{equation}
on the 1-dimensional subspace 
${\cal V_S}\cap {\cal S}_1'$
whose boundary is ${\cal S}\cap {\cal S}_1'$.
Thus, we should add this object when we try to intersect 
the 1-form symmetry generators.
Physically, the presence of the 
induced object means the production of the axion, since 
$\bs{E} \cdot \bs{B}$ becomes non-zero on
the transversal intersections of the 1-form symmetry generators, 
and $\bs{E} \cdot \bs{B}$ can be understood as a source of 
the axion.

\subsubsection{Fractional linking phase on the domain wall}
Finally, we consider the following cubic but trivial correlation 
function to show the topological order on the axionic domain wall,
\begin{equation}
\begin{split}
1& =
 \vevs{U_1 (e^{2\pi i n_1/p},{\cal S}_0)
 U_1 (e^{2\pi i n'_1/p},{\cal S}'_0) U_0 (e^{2\pi i n_0/m},{\cal V})},
 \label{eq:cubic correlation}
\end{split}
\end{equation}
where the three symmetry generators are not intersected,
${\cal S}_0 \cap {\cal S}'_0 = {\cal S}_0 \cap {\cal V} = {\cal S}'_0 \cap {\cal V} = \emptyset$.
As discussed in section~\ref{sec:Intersection of 0- and 1-form symmetry generators}, we deform the subspaces by using 
${\cal V}_{01}$ and ${\cal V}'_{01}$
satisfying $\der {\cal V}_{01} = {\cal S}_0 \cup \b{\cal S}_1$ 
and 
$\der {\cal V}'_{01} = {\cal S}'_0 \cup \b{\cal S}'_1$,
where ${\cal S}_1$ and ${\cal S}'_1$ intersect with ${\cal V}$
but ${\cal S}_1 \cap {\cal S}_1' = \emptyset$:
\begin{equation}
\begin{split}
& \vevs{U_1 (e^{2\pi i n_1/p},{\cal S}_0)
 U_1 (e^{2\pi i n'_1/p},{\cal S}'_0) U_0 (e^{2\pi i n_0/m},{\cal V})}
\\
&
\quad
= 
\langle
U_{01}(e^{2\pi i\fr{N}{mp}n_0n_1},\Omega_{{\cal V}}\cap \mathcal{S}_1)
U_{01}(e^{2\pi i\fr{N}{mp}n_0n'_1},\Omega_{{\cal V}}\cap \mathcal{S}'_1)
\\
&
\qquad
\times 
U_1 (e^{2\pi i n_1/p}, {\cal S}_1)
U_1 (e^{2\pi i n_1'/p}, {\cal S}'_1)
U_0 (e^{2\pi i n_0/m}, {\cal V}) 
\rangle.
\end{split}
\end{equation}
We then deform ${\cal S}_1'$ to ${\cal S}'_2$
by using ${\cal V}'_{12} $ whose boundaries are given as
$\der {\cal V}'_{12} = {\cal S}_1' \cup \b{\cal S}_2'$,
and ${\cal S}_2'$ intersects with ${\cal S}_1$ transversally.
The final configuration is illustrated in Fig.~\ref{U011}.
{\small
\begin{figure}[t]
\begin{center}
\ig[height=10em]{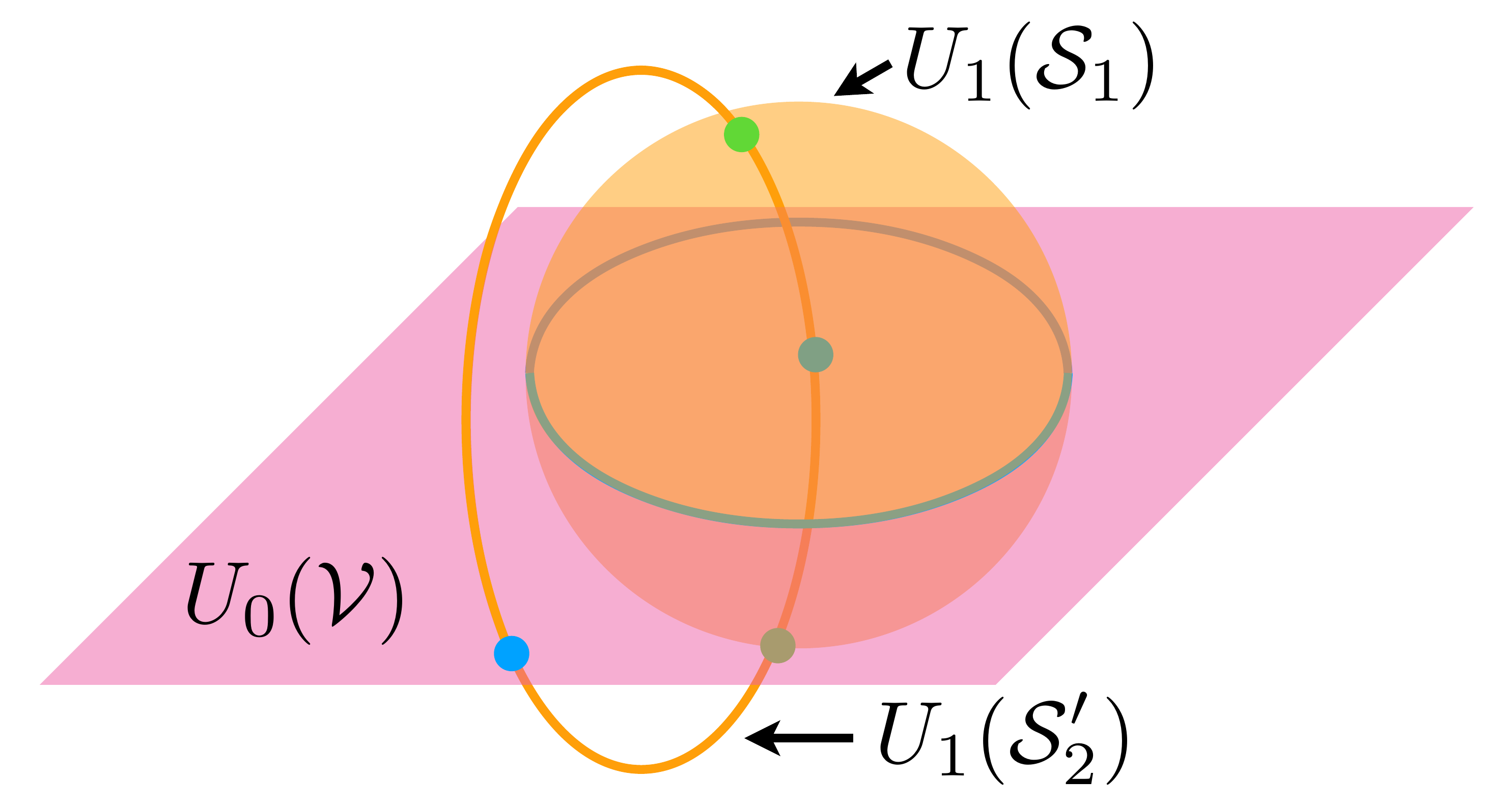}
\end{center}
\caption{\label{U011}
The intersection of a 0-form symmetry generator and
two 1-form symmetry generators at a time slice.
The configuration of the 0- and 1-form symmetry generators are the same as 
the right panel of Figs.~\ref{U01} and \ref{U11}, respectively.
The blue dots and circle represent the intersections of 
0- and 1-form symmetry generators on the time slice, respectively, and they are linked with each other
on the worldvolume ${\cal V}$.
The green dots mean the intersections of 
the 1-form symmetry generators.
We have omitted the induced objects $U_{01}$ 
 and $U_{11}$ to avoid the complication of the figure.
They exist such that their boundaries are the blue dots and blue circle
for $U_{01}$ and green dots for $U_{11}$.
We have again abbreviated the parameters of the symmetry generators.
}
\end{figure}
}
By the deformation, we have
\begin{equation}
\begin{split}
& \vevs{U_1 (e^{2\pi i n_1/p},{\cal S}_0)
 U_1 (e^{2\pi i n'_1/p},{\cal S}'_0) U_0 (e^{2\pi i n_0/m},{\cal V})}
\\
&
\quad= 
\langle
U_{01}(e^{2\pi i\fr{N}{mp}n_0n_1},\Omega_{{\cal V}}\cap \mathcal{S}_1)
U_{01}(e^{2\pi i\fr{N}{mp}n_0n'_1},\Omega_{{\cal V}}\cap \mathcal{S}'_1)
\\
&
\qquad
\times 
U_1 (e^{2\pi i n_1/p}, {\cal S}_1)
U_1 (e^{2\pi i n_1'/p}, \der{\cal V}'_{12})
U_1 (e^{2\pi i n_1'/p}, {\cal S}'_2)
U_0 (e^{2\pi i n_0/m}, {\cal V}) 
\rangle.
\end{split}
\end{equation}
By the redefinition 
$a + \fr{2\pi n'_1}{p} \delta_1 ({\cal V}'_{12}) \to a$,
we obtain induced objects $U_{01}$ and $U_{11} $
as well as constant phases:
\begin{equation}
\begin{split}
& \vevs{U_1 (e^{2\pi i n_1/p},{\cal S}_0)
 U_1 (e^{2\pi i n'_1/p},{\cal S}'_0) U_0 (e^{2\pi i n_0/m},{\cal V})}
\\
&
= 
e^{- 2\pi i\fr{N}{mp^2}n_0 n_1n_1' 
\int_{\Omega_{{\cal V}}} d\delta_1 ({\cal V}'_{12}) \wed 
d\delta_1 ({\cal V}_{01})
}
e^{- 2\pi i\fr{N}{mp^2}n_0n'^2_1  
 \int_{\Omega_{{\cal V}}} 
d
 \delta_1 ({\cal V}'_{12}) 
 \wed 
d\delta_1 ({\cal V}'_{01})}
\\
&
\quad
\times 
\langle
U_{01}(e^{2\pi i\fr{N}{mp}n_0n_1},\Omega_{{\cal V}}\cap \mathcal{S}_1)
U_{01}(e^{2\pi i\fr{N}{mp}n_0n'_1},\Omega_{{\cal V}}\cap \mathcal{S}'_2)
\\
&
\quad\qquad
\times 
U_{11}(e^{2\pi i \fr{N}{p^2} n_1n_1'},\mathcal{V}_{\mathcal{S}_1}\cap\mathcal{S}'_2 )
\\
&
\quad\qquad
\times 
U_1 (e^{2\pi i n_1/p}, {\cal S}_1)
U_1 (e^{2\pi i n_1'/p}, {\cal S}'_2)
U_0 (e^{2\pi i n_0/m}, {\cal V}) 
\rangle.
\end{split}
\end{equation}
Due to the relations 
\begin{equation}
\begin{split}
 \int_{\Omega_{{\cal V}}} d\delta_1 ({\cal V}'_{12}) \wed 
d\delta_1 ({\cal V}_{01})
&= 
 \int_{\Omega_{\cal V}}
\delta_2({\cal S}'_2) 
\wed 
\delta_2({\cal S}_1) 
= 
- \int_{{\cal V}}
\delta_2({\cal S}_1) 
\wed 
\delta_1({\cal V}_{{\cal S}'_2}) 
\\
&=: 
-\link ({\cal S}_1, {\cal S}'_2)|_{\cal V}, 
\end{split}
\end{equation}
\begin{equation}
  \int_{\Omega_{{\cal V}}} 
d
 \delta_1 ({\cal V}'_{12}) 
 \wed 
d\delta_1 ({\cal V}'_{01})
=
 \int_{\Omega_{{\cal V}}} 
( \delta_2 ({\cal S}'_1) -  \delta_2 ({\cal S}'_2))
 \wed 
( \delta_2 ({\cal S}'_0) -  \delta_2 ({\cal S}'_1)) =0,
\end{equation}
we find 
\begin{equation}
\begin{split}
 &\vevs{U_1 (e^{2\pi i n_1/p},{\cal S}_0)
 U_1 (e^{2\pi i n'_1/p},{\cal S}'_0) U_0 (e^{2\pi i n_0/m},{\cal V})}
\\
&
\quad= 
e^{
2\pi i\fr{N}{mp^2}n_0 n_1n_1' 
\link ({\cal S}_1, {\cal S}'_2)|_{\cal V}
}
\\
&
\qquad
\times 
\langle
U_{01}(e^{2\pi i\fr{N}{mp}n_0n_1},\Omega_{{\cal V}}\cap \mathcal{S}_1)
U_{01}(e^{2\pi i\fr{N}{mp}n_0n'_1},\Omega_{{\cal V}}\cap \mathcal{S}'_2)
U_{11}(e^{2\pi i \fr{N}{p^2} n_1n_1'},\mathcal{V}_{\mathcal{S}_1}\cap\mathcal{S}'_2 )
\\
&
\qquad\qquad
\times 
U_1 (e^{2\pi i n_1/p}, {\cal S}_1)
U_1 (e^{2\pi i n_1'/p}, {\cal S}'_2)
U_0 (e^{2\pi i n_0/m}, {\cal V}) 
\rangle.
\end{split}
\end{equation}
Here, ${\cal V}_{{\cal S}'_2}$ is 
a 3-dimensional subspace whose boundary is ${\cal S}'_2$, 
and
the symbol ``$\link ({\cal S}_1, {\cal S}'_2)|_{\cal V}$''
is a linking number of ${\cal S}_1$ and ${\cal S}'_2$
on the closed 3-dimensional subspace ${\cal V}$.
In other words, using Eq.~\eqref{eq:cubic correlation}, we have 
\begin{equation}
\begin{split}
&
\langle
U_{01}(e^{2\pi i\fr{N}{mp}n_0n_1},\Omega_{{\cal V}}\cap \mathcal{S}_1)
U_{01}(e^{2\pi i\fr{N}{mp}n_0n'_1},\Omega_{{\cal V}}\cap \mathcal{S}'_2)
U_{11}(e^{2\pi i \fr{N}{p^2} n_1n_1'},\mathcal{V}_{\mathcal{S}_1}\cap\mathcal{S}'_2 )
\\
&
\qquad
\times 
U_1 (e^{2\pi i n_1/p}, {\cal S}_1)
U_1 (e^{2\pi i n_1'/p}, {\cal S}'_2)
U_0 (e^{2\pi i n_0/m}, {\cal V}) 
\rangle
\\
&
= 
e^{ -
2\pi i\fr{N}{mp^2}n_0 n_1n_1' 
\link ({\cal S}_1, {\cal S}'_2)|_{\cal V}
}
.
\end{split}
\end{equation}
The final form implies that the anyons on the domain wall induced by
magnetic fluxes have a fractional linking phase, which implies the
topological order.

\section{\label{sum}Summary and Discussion}

In this paper, we have investigated the higher-form symmetries 
in the topological axion electrodynamics in $(3+1)$ dimensions.
We have coupled 
the background gauge fields for 
0-, 1-, 2-, and 3-form symmetries to the action.
By the gauge invariance for the 
axion and photon, 
we have found that the gauging of the 1-form symmetry 
requires the simultaneous gauging of the 3-form symmetry, 
and the simultaneous gauging of the 0- and 1-form symmetries 
requires the gauging of the 2-form symmetry.
These requirements modify the fractional AB phases of 
the background gauge fields for the 2- and 3-form symmetries.
By these modifications, we have found that 
the groups of the higher-form symmetries organize the semi-strict 4-group
or 3-crossed module.

We further have derived 't Hooft anomalies of the 4-group symmetry.
There are mixed 't Hooft anomalies between the 0- and 3-form symmetries as well as the 1- and 2-form symmetries.
Furthermore, we have found a mixed 't Hooft anomaly between 
the 0- and 
1-form symmetries, which is a 2-group anomaly.
We have then discussed physical consequences derived by the 't Hooft anomalies.
In particular, we have shown the topological order on the axionic domain walls by using the 2-group anomaly.
We have also given a detailed derivation of the topological order in terms of the symmetry generators with a careful treatment of the intersections of symmetry generators.

There are several avenues for future work.
We can develop mathematical foundations 
of 4-group gauge theories based on the semi-strict 4-group.
It would be a nontrivial question how we can treat 
several types of the Peiffer lifting proposed in Ref.~\cite{Arvasi:2009}
to construct the gauge theories.
In particular, the 3-crossed module may have other types 
of a Peiffer lifting such as $G_1\times G_2\to G_3 $~\cite{Arvasi:2009}.
We expect that they also express the presence of 
boundaries of symmetry generators on intersections 
of symmetry generators.

To apply the semi-strict 4-group to physics,
it would be useful to understand the 4-group diagrammatically.
We may extend 
a diagrammatic expression of the semi-strict 3-group proposed in Ref.~\cite{Hidaka:2020izy}
to the 4-group.
It is also possible to apply our framework to 
a low-energy effective theory of topological
superconductors 
in $(3+1)$ dimensions, since 
they can be described by the massive photon and axions 
with topological couplings between them~\cite{Qi:2012cs,Stone:2016pof,Stalhammar:2021tcq}.

\subsection*{Acknowledgements}
RY thanks Ryohei Kobayashi, Tatsuki Nakajima, Tadakatsu Sakai, and Yuya Tanizaki 
for helpful discussions. 
This work is supported in part by Japan Society of Promotion of Science (JSPS) Grant-in-Aid for 
Scientific Research (KAKENHI Grants No.~JP17H06462, JP18H01211 (YH), JP18H01217 (MN), 
JP21J00480, JP21K13928 (RY)).

\appendix
\section{Derivations of symmetry transformations \label{corr}}

We have shown the symmetry transformations 
of higher-form symmetries in terms of correlation functions
 in \ers{210516.1529-0}--\eqref{210516.1529-3}.
Here, we summarize the derivations of the symmetry transformations.
The derivations are based on reparameterizations of dynamical 
fields, 
which are finite versions of Schwinger-Dyson equations.

Before showing the derivations, it will be convenient to 
denote 0-,..., 3-form symmetry generators by using conserved currents 
$j_3$,..., $j_0$ as follows,
\begin{align}
U_0 (e^{2\pi i n_0/m }, {\cal V})
 &
=
e^{\fr{2\pi i n_0}{m} \int_{\cal V} j_3}  ,
\\
U_1 (e^{2\pi i n_1/p }, {\cal S})
 &
=
e^{\fr{2\pi i n_1}{p} \int_{\cal S} j_2}  ,
\\
U_2 (e^{2\pi i n_2/q }, {\cal C})
 &
=
e^{\fr{2\pi i n_2}{q} \int_{\cal C} j_1}  ,
\\
U_3 (e^{2\pi i n_3/k }, ({\cal P,P'}))
 &
=
e^{\fr{2\pi i n_3}{k}  (j_0 ({\cal P})- j_0 ({\cal P'}))} , 
\end{align}
where 
the integrands of the 0-,..., 3-form symmetry generators are
\begin{align}
j_3 &=  -\fr{k}{2\pi} c - \fr{N}{8\pi^2} a \wed da,
\\
j_2 & =  - \fr{q}{2\pi} b - \fr{N}{4\pi^2} \phi da,
\\
 j_1 & = -\fr{q}{2\pi} a, 
\\
 j_0 & = -\fr{k}{2\pi} \phi,
\end{align}
respectively.
These currents are closed $dj_i=0$ by the equations of motion.
Note that the subscripts of the currents indicate the degrees of the differential forms.

\subsection{$\bb{Z}_m$ 0-form symmetry}
Now, we derive the symmetry transformations.
First, we consider the symmetry transformation of the 0-form symmetry.
In the path-integral formalism,
the left-hand side of \er{210516.1529-0} can be expressed
as 
\begin{equation}
 \vevs{  U_0 (e^{2\pi i n_0/m},{\cal V}) L (q_0, {\cal P}) }
= {\cal N} 
\int {\cal D} [\phi, a,b, c] 
e^{iS_{\rm TAE} [\phi, a,b, c] + \fr{2\pi i n_0}{m} \int_{\cal V} j_3
+ iq_0 \phi ({\cal P})}.
\end{equation}
Here, 
${\cal N}$ is the normalization factor so that $\vevs{1} =1$,
and the symbol ``${\cal D} [\phi, a,b, c] $''
stands for the integral measure 
${\cal D} \phi {\cal D} a {\cal D}b {\cal D} c $.

The correlation function can be evaluated by absorbing 
the symmetry generator to the action.
To see this, 
we express $\int_{\cal V} j_3$ by using the Stokes theorem
\begin{equation}
 \int_{\cal V} j_3 
= \int_{\Omega_{\cal V}} dj_3
= \int_{M_4} dj_3 \wed \delta_0(\Omega_{\cal V})
\end{equation}
for a 4-dimensional space $\Omega_{\cal V}$ satisfying 
$\der \Omega_{\cal V} = {\cal V}$.
By using the relation,
\begin{equation}
 S_{\rm TAE}
\left[\phi - \fr{2\pi n_0 }{m} \delta_0 (\Omega_{\cal V}),a,b,c \right]
= 
 S_{\rm TAE}[\phi, a,b,c]
+
\fr{2\pi n_0 }{m} 
\int_{M_4} dj_3 \wed \delta_0(\Omega_{\cal V}),
\end{equation}
we can absorb the symmetry generator into the action as
\begin{equation}
 \vevs{  U_0 (e^{2\pi i n_0/m},{\cal V}) L (q_0, {\cal P}) }
= {\cal N} 
\int {\cal D} [\phi, a,b, c] 
e^{iS_{\rm TAE} \left[\phi - \fr{2\pi n_0 }{m} \delta_0 (\Omega_{\cal V}),a,b,c \right]
+ iq_0 \phi ({\cal P})}.
\end{equation}
By the reparameterization 
$\phi - \fr{2\pi n_0 }{m} \delta_0 (\Omega_{\cal V}) \to \phi$, 
we obtain the relation in \er{210516.1529-0}:
\begin{equation}
 \vevs{  U_0 (e^{2\pi i n_0/m},{\cal V}) L (q_0, {\cal P}) }
=
e^{i \fr{2\pi q_0 n_0}{m} \link ({\cal V, P})}
\vevs{L (q_0, {\cal P})}.
\end{equation}
Here, we have used 
$\phi ({\cal P}) = \int_{M_4} \phi (x) \delta_4 ({\cal P})$
with
$\delta_4 ({\cal P}) = \delta^4 (x- {\cal P}) dx^0 \weds dx^3$,
and
the definition of the linking number,
\begin{equation}
\link ({\cal V, P})
= \int_{\Omega_{\cal V}} \delta_4 ({\cal P}). 
\end{equation}

\subsection{$\bb{Z}_p$ 1-form symmetry}
Second, we study the 1-form symmetry transformation 
in \er{210516.1529-1}.
The correlation function in \er{210516.1529-1}
can be expressed as
\begin{equation}
 \vevs{  U_1 (e^{2\pi i n_1/p},{\cal S}) W (q_1, {\cal C}) }
= {\cal N} 
\int {\cal D} [\phi, a,b, c] 
e^{iS_{\rm TAE} [\phi, a,b, c] + \fr{2\pi i n_1}{p} \int_{\cal S} j_2
+ i q_1 \int_{\cal C} a}.
\end{equation}
As in the case of the 0-form symmetry transformation, 
we absorb the symmetry generator to the action.
By using the Stokes theorem,
\begin{equation}
 \int_{\cal S} j_2 
= \int_{{\cal V_S}} dj_2 = \int_{M_4} dj_2 \wed \delta_1 ({\cal V_S}),
\end{equation}
and using the relation,
\begin{equation}
\begin{split}
& S_{\rm TAE} 
\left[\phi, a + \fr{2\pi  n_1}{p} \delta_1 ({\cal V_S}),b,c \right]
\\
&= 
 S_{\rm TAE} [\phi, a, b, c]
+ \frac{2\pi n_1}{p}\int_{M_4} dj_2 \wed \delta_1 ({\cal V_S})
+ \fr{N}{8\pi^2}\cdot \(\fr{2\pi  n_1}{p} \)^2
\int_{M_4} \phi \delta_2 ({\cal S}) \wed  \delta_2 ({\cal S}),
\end{split}
\end{equation}
the correlation function can be written as
\begin{equation}
 \vevs{  U_1 (e^{2\pi i n_1/p},{\cal S}) W (q_1, {\cal C}) }
= {\cal N} 
\int {\cal D} [\phi, a,b, c] 
e^{iS_{\rm TAE} 
\left[\phi, a + \fr{2\pi  n_1}{p} \delta_1 ({\cal V_S}),b,c \right]
+ i q_1 \int_{\cal C} a}.
\end{equation}
Here, we have used the assumption that ${\cal S}$ does not 
have self-intersections, 
$\delta_2 ({\cal S}) \wed  \delta_2 ({\cal S}) =0$.
By the reparameterization 
$a + \fr{2\pi  n_1}{p} \delta_1 ({\cal V_S}) \to a$,
we arrive at 
\begin{equation}
  \vevs{  U_1 (e^{2\pi i n_1/p},{\cal S}) W (q_1, {\cal C}) }
= 
e^{\fr{2\pi i q_1 n_1 }{p } \link ({\cal S,C})}
 \vevs{  W (q_1, {\cal C}) },
\end{equation}
where we have used
\begin{equation}
 \link ({\cal S, C})
= \int_{{\cal V_S}} \delta_3 ({\cal C})
=- \int_{M_4} \delta_1 ({\cal V_S}) \wed  \delta_3 ({\cal C}). 
\end{equation}

\subsection{$\bb{Z}_q$ 2-form symmetry}
Third, we discuss the 2-form symmetry transformation 
in \er{210516.1529-2}.
The derivation is similar to those of 0- and 1-form symmetry 
transformations as we discussed above.
The correlation function in \er{210516.1529-2}
can be written as 
\begin{equation}
 \vevs{  U_2 (e^{2\pi i n_2/q},{\cal C}) V (q_2, {\cal S}) }
= {\cal N} 
\int {\cal D} [\phi, a,b, c] 
e^{iS_{\rm TAE} [\phi, a,b, c] + \fr{2\pi i n_2}{q} \int_{\cal C} j_1
+ i q_2 \int_{\cal S} b}.
\end{equation}
By using the Stokes theorem
\begin{equation}
 \int_{\cal C} j_1 = \int_{\cal S_C} dj_1 =
\int_{M_4} dj_1 \wed \delta_2 ({\cal S_C})
\end{equation}
and the relation
\begin{equation}
 S_{\rm TAE} \left[ \phi, a, b - \fr{2\pi n_2}{q} \delta_2({\cal S_C}), c\right]
=  S_{\rm TAE} \left[ \phi, a, b, c\right]
 + \fr{2\pi n_2}{q}\int dj_1 \wed \delta_2({\cal S_C}),
\end{equation}
we obtain 
\begin{equation}
 \vevs{  U_2 (e^{2\pi i n_2/q},{\cal C}) V (q_2, {\cal S}) }
= {\cal N} 
\int {\cal D} [\phi, a,b, c] 
e^{i S_{\rm TAE} \left[ \phi, a, b - \fr{2\pi n_2}{q} \delta_2({\cal S_C}), c\right]
+ i q_2 \int_{\cal S} b}.
\end{equation}
By the reparameterization 
$b - \fr{2\pi n_2}{q} \delta_2({\cal S_C}) \to b$,
we obtain 
\begin{equation}
 \vevs{  U_2 (e^{2\pi i n_2/q},{\cal C}) V (q_2, {\cal S}) }
= 
e^{\fr{2\pi i q_2 n_2}{q} \link ({\cal C,S})}
 \vevs{  V (q_2, {\cal S}) },
\end{equation}
where we have used
\begin{equation}
 \link ({\cal C,S}) 
= 
\int_{\cal S_C} \delta_2 ({\cal S})
= 
\int \delta_2 ({\cal S_C}) \wed \delta_2 ({\cal S}) .
\end{equation}
\subsection{$\bb{Z}_k$ 3-form symmetry}
Finally, 
we consider the 3-form symmetry transformation 
in \er{210516.1529-3}.
The correlation function in \er{210516.1529-3}
can be written as 
\begin{equation}
 \vevs{  U_3 (e^{2\pi i n_3/q},({\cal P,P'})) D (q_3, {\cal V}) }
= {\cal N} 
\int {\cal D} [\phi, a,b, c] 
e^{iS_{\rm TAE} [\phi, a,b, c] + \fr{2\pi i n_3}{k} 
(j_0 ({\cal P}) - j_0 ({\cal P'}))
+ i q_3\int_{\cal V} c}.
\end{equation}
By using the line integral on a 1-dimensional subspace 
${\cal C_{P,P'}}$ satisfying 
$\der {\cal C_{P,P'}} = {\cal P }\cup \b{\cal P}'$
\begin{equation}
 j_0 ({\cal P}) - j_0 ({\cal P'})
 = \int_{\cal C_{P,P'}} dj_0 
=
\int_{M_4} dj_0 \wed \delta_3 ({\cal C_{P,P'}}),
\end{equation}
and the relation
\begin{equation}
 S_{\rm TAE} \left[ \phi, a, b , c + \fr{2\pi n_3}{k} \delta_3 ({\cal C_{P,P'}})\right]
=  S_{\rm TAE} \left[ \phi, a, b, c\right]
 + \fr{2\pi n_3}{k}\int_{M_4} dj_0 \wed \delta_3 ({\cal C_{P,P'}}),
\end{equation}
we obtain 
\begin{equation}
 \vevs{  U_3 (e^{2\pi i n_3/q},({\cal P,P'})) D (q_3, {\cal V}) }
= {\cal N} 
\int {\cal D} [\phi, a,b, c] 
e^{iS_{\rm TAE} \left[ \phi, a, b , c + \fr{2\pi n_3}{k} \delta_3 ({\cal C_{P,P'}})\right]
+ i q_3\int_{\cal V} c}.
\end{equation}
By the reparameterization 
$c + \fr{2\pi n_3}{k} \delta_3 ({\cal C_{P,P'}})  \to c$,
we obtain 
\begin{equation}
  \vevs{  U_3 (e^{2\pi i n_3/q},({\cal P,P'})) D (q_3, {\cal V}) }
= 
e^{\fr{2\pi i q_3 n_3}{k} \link ({\cal (P,P'),V})}
 \vevs{ D (q_3, {\cal V}) },
\end{equation}
where we have used
\begin{equation}
 \link ({\cal (P,P'),V})
= 
\int_{\cal C_{P,P'}} \delta_1 ({\cal V})
= 
- \int \delta_3 ({\cal C_{P,P'}}) \wed \delta_1 ({\cal V}) .
\end{equation}

\providecommand{\href}[2]{#2}\end{document}